\documentclass[letterpaper, paper,11pt]{AAS}	

\usepackage{bm}
\usepackage{amsmath}
\usepackage{subfigure}
\usepackage{subcaption} 

\usepackage[colorlinks=true, pdfstartview=FitV, linkcolor=black, citecolor= black, urlcolor= black]{hyperref}
\usepackage{overcite}
\usepackage{footnpag}			      	
\usepackage{hyperref}
\usepackage{footmisc}

\usepackage{tabularx}
\usepackage{longtable}
\usepackage{graphicx}
\usepackage{float}

\PaperNumber{25-394}

\begin{document}

\newcommand{\affilA}{\footnotemark[1]} 
\newcommand{\affilB}{\footnotemark[2]} 
\newcommand{\affilC}{\footnotemark[3]} 
\newcommand{\affilD}{\footnotemark[4]} 
\newcommand{\affilE}{\footnotemark[5]} 

\title{MILANI AND JUVENTAS OPTICAL NAVIGATION IMAGES IN SUPPORT OF THE HERA RADIO SCIENCE EXPERIMENT }
\author{Edoardo Gramigna\affilA,
Ozgur Karatekin\affilB,
Birgit Ritter\affilB,
Iosto Fodde\affilC,\\
Fabio Ferrari\affilC,
Felice Piccolo\affilC,
Paolo Tortora\affilA\,\affilD,
Riccardo Lasagni Manghi\affilD,
Marco Zannoni\affilA\,\affilD,
Ryan S. Park\affilE
}

\maketitle{} 		
\footnotetext[1]{Centro Interdipartimentale di Ricerca Industriale Aerospaziale (CIRI AERO) Alma Mater Studiorum – Università di Bologna, Via Baldassarre Carnaccini 12, 47121 Forlì, Italy. edoardo.gramigna@unibo.it}

\footnotetext[2]{Royal Observatory of Belgium, Brussels, Belgium} 

\footnotetext[3]{Department of Aerospace Science and Technology (DAER), Politecnico di Milano, Via La Masa 34, Milan, Italy} 

\footnotetext[4]{Department of Industrial Engineering (DIN), University of Bologna, via Fontanelle 40, 47121, Forlì, Italy.} 

\footnotetext[5]{NASA Jet Propulsion Laboratory, California Institute of Technology, Pasadena, CA, USA.}

\begin{abstract}
The European Space Agency's Hera mission aims to advance planetary defense by investigating the Didymos binary asteroid system following NASA's DART impact on Dimorphos. Incorporating simulated CubeSat images into Hera's orbit determination framework significantly improves positional accuracy, reducing uncertainties by up to three orders of magnitude for Milani and one for Juventas. This data enable more precise estimation of key parameters such as asteroid masses, rotational states, and moments of inertia, with up to 90\% uncertainty reduction in Dimorphos’ dynamical state under chaotic scenarios. Sensitivity analysis reveals the influence of pointing accuracy, image noise, and imaging frequency. Additionally, dedicated scientific investigations using Dimorphos-focused CubeSat images further advance understanding of the system. These findings underscore the critical role of CubeSat images in refining navigation and advancing science, offering valuable insights for future small-body missions.

\end{abstract}

\section{Introduction}

Hera is the European Space Agency’s inaugural planetary defense space mission \cite{michel2022hera} and represents the European counterpart to the Asteroid Impact and Deflection Assessment (AIDA) international collaboration \cite{cheng2018aida}, alongside NASA’s DART mission. The first asteroid deflection experiment was conducted on September 26, 2022, when NASA's DART mission intentionally impacted Dimorphos, the small moon of the Didymos binary asteroid system, using a kinetic impactor technique \cite{cheng2023momentum}. In this context, Hera's goal is to comprehensively assess and characterize the feasibility of the kinetic impactor technique for asteroid deflection, while also conducting an in-depth investigation of the asteroid binary. This includes examining its physical and compositional properties, as well as studying the effect of the impact on the surface and shape of Dimorphos.

Hera is accompanied by two companion 6U CubeSats, Milani and Juventas, which will be released in the vicinity of the Didymos system \cite{goldberg2019juventas,karatekin2021juventas,ferrari2021preliminary,ferrari2021trajectory}. The CubeSats are tracked by Hera using an Inter-Satellite Link (ISL) system, the first miniaturized satellite-to-satellite transceiver ever flown in deep space. This is a crucial addition to the gravity investigation, as the relative Doppler shift between Hera and the CubeSats, measured via the ISL, provides highly valuable information on the system’s dynamics. This is due to the closer proximity of the CubeSats to the asteroids, as demonstrated by Gramigna et al. (2024) \cite{gramigna2024hera}.

The Hera Radio Science Experiment (RSE) \cite{gramigna2024hera} utilizes precise orbit determination to estimate key physical parameters necessary for calculating the momentum enhancement resulting from the DART impact and conducting a post-impact survey of the system. These parameters include the mass, mass distribution, rotational states, relative orbits, and dynamics of the Didymos and Dimorphos asteroids. Indeed, the gravity, orbits, and rotational states of Didymos and Dimorphos can be determined by accurately reconstructing the trajectories of the Hera, Juventas, and Milani spacecraft during the nominal asteroid phase \cite{gramigna2022hera}. The expected accuracies of the experiment are summarized in Gramigna et al. (2024) \cite{gramigna2024hera, gramigna2024heraAAS}.

The main measurements used in the nominal Orbit Determination (OD) process include Hera's two-way Doppler and range observables at X-band, obtained through the standard Earth-based communication link; ISL two-way range observables between Hera and the CubeSats at S-band; and optical navigation (opnav) images of Didymos and Dimorphos captured by Hera’s Asteroid Framing Camera (AFC).

This work aims to enhance the OD framework of the Hera mission by simulating the opnav images expected from the CubeSats Juventas and Milani during the asteroid phase. Hera is the first deep-space mission to undertake such a complex investigation, leveraging data from multiple instruments and spacecraft. The optical measurements acquired by Milani and Juventas enable improved triangulation of the spacecraft positions within the Didymos system, augmenting the performance achievable using Earth-based tracking, ISL data, and Hera's onboard images.

The primary objective of this research is to characterize and quantify the improvement in estimating key scientific parameters facilitated by these additional observables, compared to the baseline scenario that excludes CubeSat images. The CubeSats' optical data significantly enhance the determination of Juventas and Milani's relative positions with respect to the asteroids, further constraining the ISL radiometric tracking and improving the OD accuracy. This refined OD, combined with the asteroids' optical data extracted from CubeSat opnav, enables more precise estimation of critical parameters, such as Dimorphos' relative orbit, the asteroids' masses, and their rotational states.

In this study, we evaluate the contributions of these additional images in both non-chaotic and chaotic dynamical regimes of the asteroids to understand the impact of CubeSat imagery under various scenarios that Hera, Juventas, and Milani may encounter. Additionally, we conduct parametric analyses to assess the sensitivity of the results to different CubeSat operational conditions and explore the scientific potential of the findings to inform the design of Juventas and Milani's optical investigations.


\section{Juventas and Milani CubeSats}

Juventas CubeSat is developed by a consortium led by GomSpace and devoted to the geophysical characterization of Dimorphos. The main objectives are to determine the extended gravity field of Dimorphos, as well as the interior structure and surface properties of Dimorphos.

\begin{figure}[htbp]
    \centering
    \includegraphics[width=0.3\textwidth]{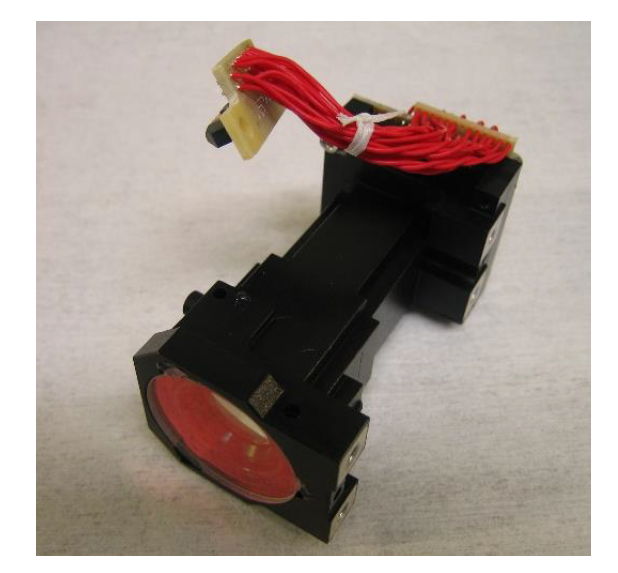}
    \caption{Juventas navigation camera.}
    \label{figjuv1}
\end{figure}

\begin{figure}[htbp]
    \centering
    \includegraphics[width=\textwidth]{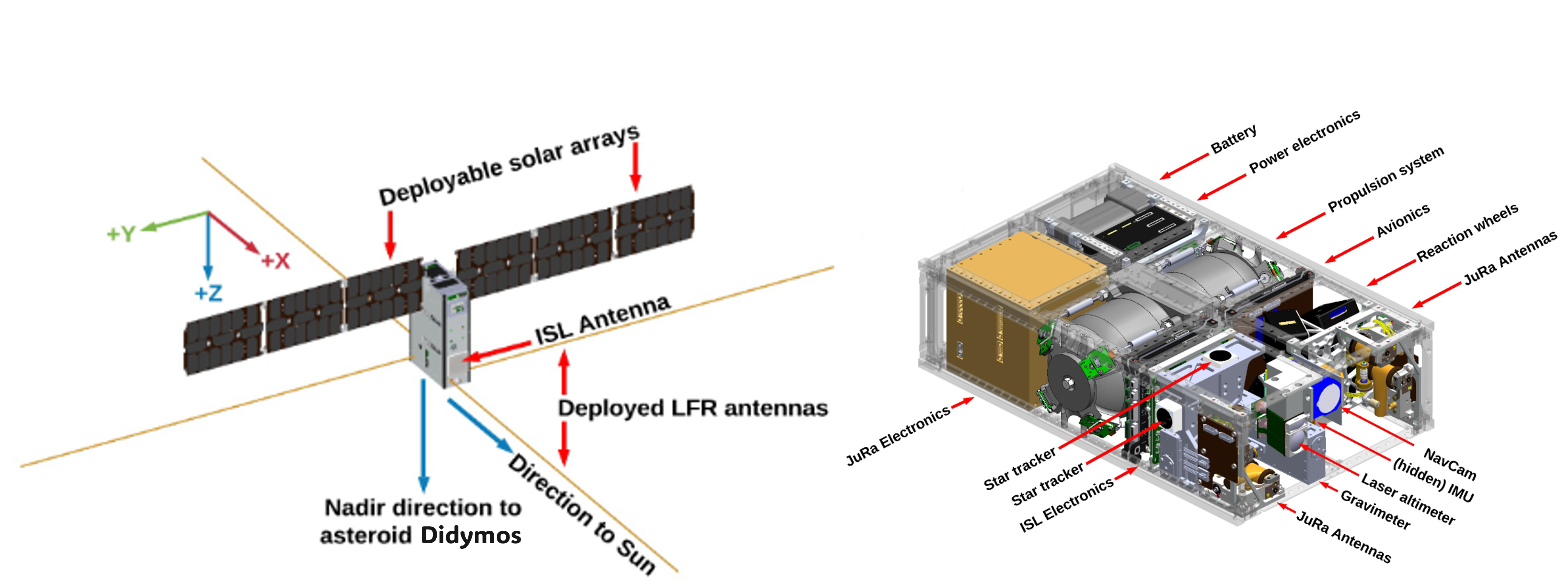}
    \caption{Juventas CubeSat (left) and its instruments (right). Image credit: GOMSpace.}
    \label{figjuv2}
\end{figure}

The Juventas navigation camera (NavCam) is used primarily for navigation purposes. It is a 58x36x33 mm visible imager with a mass of 180 g, and placed in a co-boresight configuration with the laser altimeter instrument. The resolution of the camera is 2048x1536 pixel, with a Field Of View (FOV) of 38.0x28.5 deg, monochrome channel and update rate of 2 Hz, see Figures \ref{figjuv1}-\ref{figjuv2}. A rectangular crop is applied to remove the dark and uninteresting pixels. Additionally, 256 x 256 pixel thumbnail images are constructed from the full-size images. These can be downlinked more frequently and used on ground to decide which full-size images to downlink. 

\begin{figure}[htbp]
	\centering
	\includegraphics[width=0.6\textwidth]{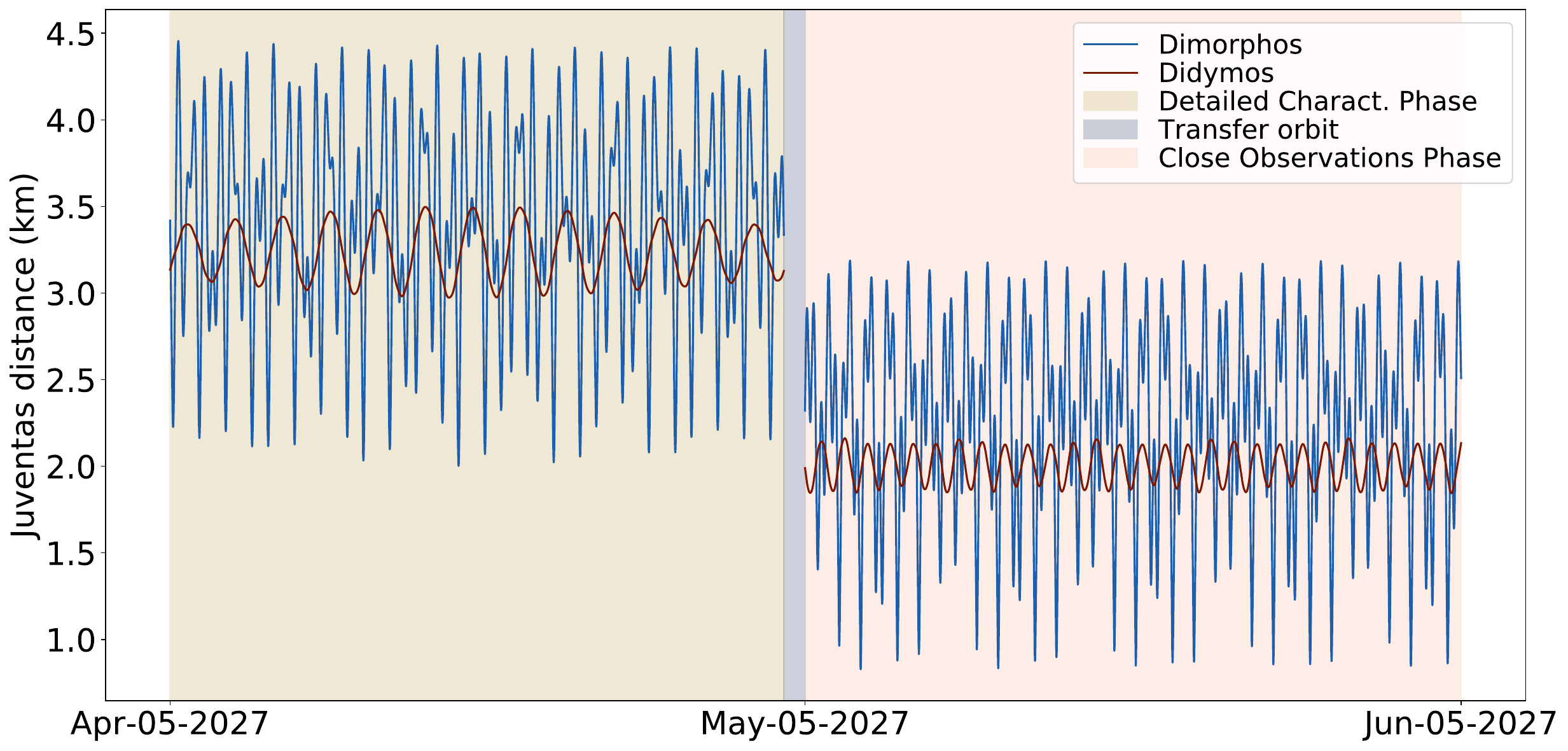}
	\caption{Juventas distances to Didymos and Dimorphos during the two SSTO orbits around Didymos at 3.3 km and 2.0 km altitude.}
	\label{figjuv}
\end{figure}

The images obtained by Juventas NavCam during its Self Stabilized Terminator Orbits (SSTO), which are reported in Figure \ref{figjuv}, as well as landing and surface operations are extremely important for science, too. In particular they allow for shape models reconstruction (together with laser ranging), surface geology, potential changes in surface, track any debris in the system, rotational and dynamical state estimation (in combination with RSE and laser altimeter), DART impact site characterization, landing site selection, and attitude determination during surface operations. In this context, Dimorphos images are not as effective for navigation purposes, but can be taken for scientific purposes and is investigated at the end of the Results Section.


\begin{figure}[htbp]
    \centering
    \includegraphics[width=\textwidth]{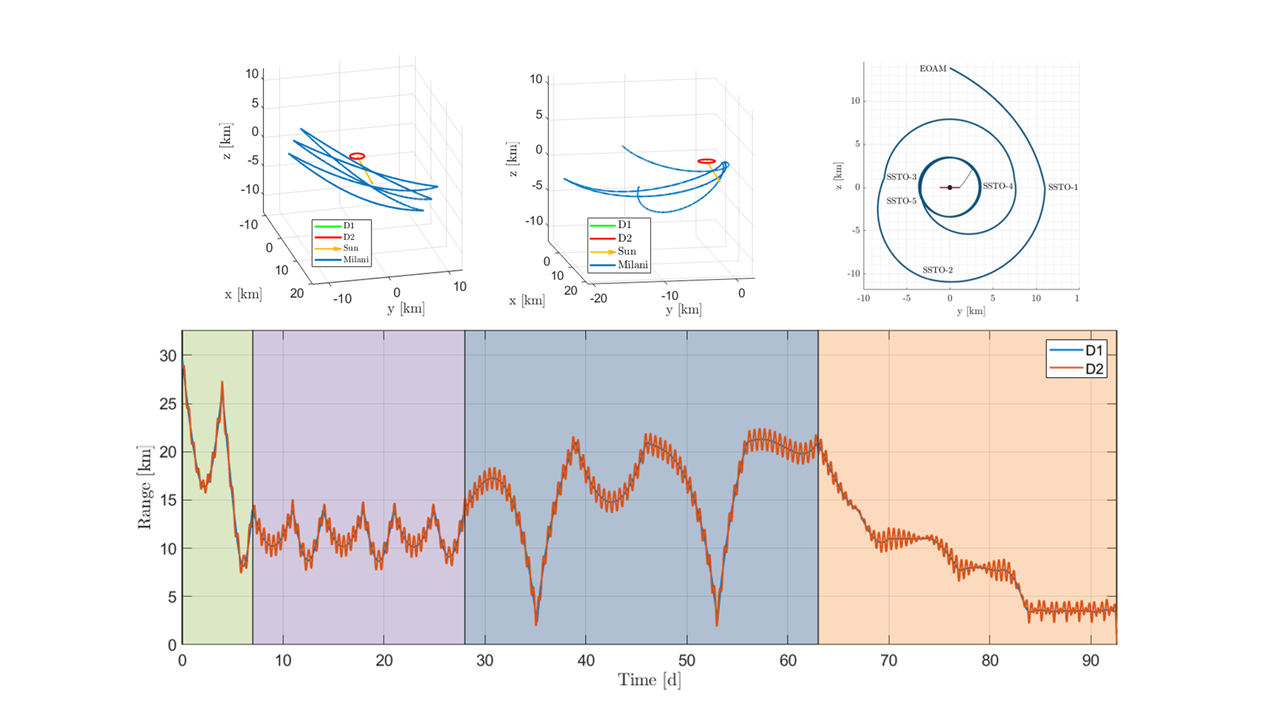}
    \caption{The different phases of Milani, where D1 is Didymos and D2 is Dimorphos.}
    \label{fig:milani_phases}
\end{figure}
\begin{figure}[htbp]
    \centering
    \includegraphics[width=0.4\textwidth]{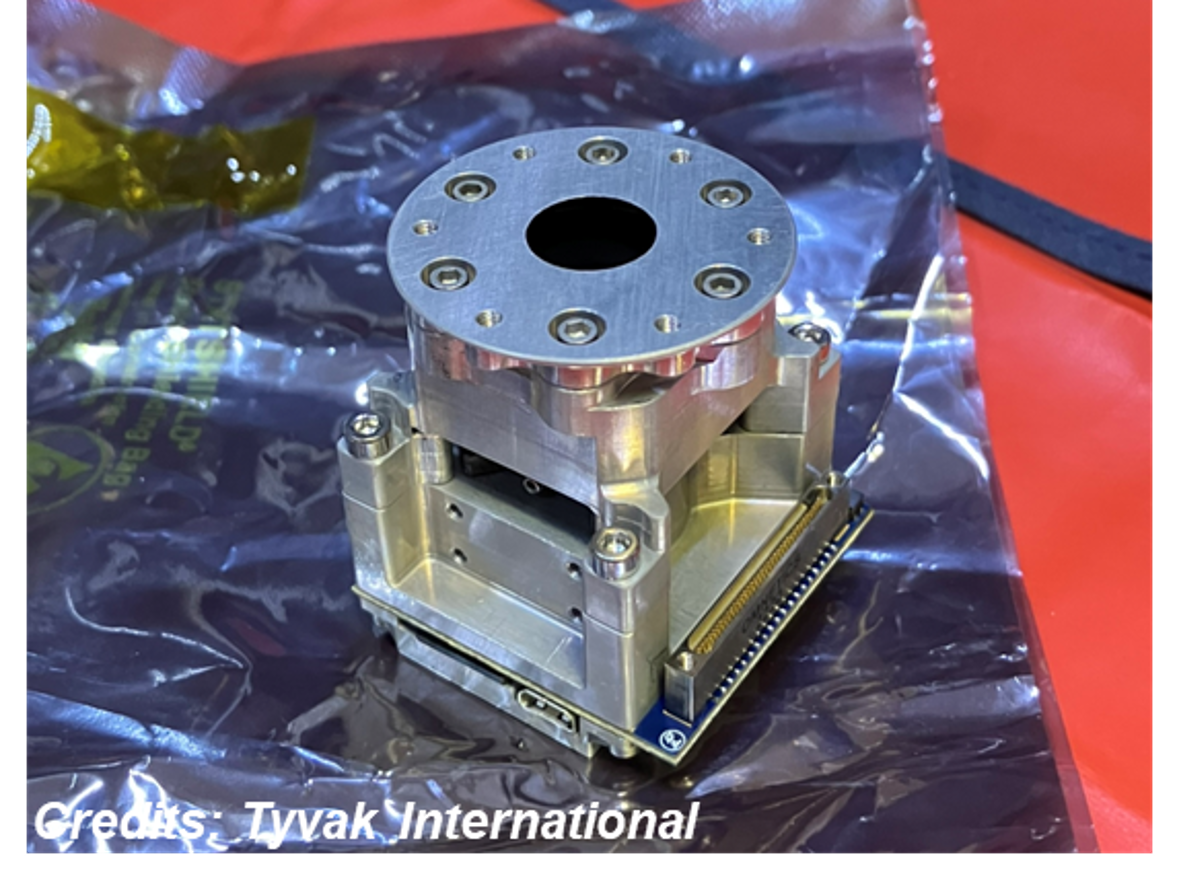}
    \caption{The Milani NavCam with baffle. Image credit: Tyvak.}
    \label{fig:milani_navcam}
\end{figure}

Milani's CubeSat main scientific objectives are: to characterize the surface and dynamical environment of both bodies, investigate the dust environment around the system, and provide measurements for supporting the radio science experiment. Besides the scientific objectives, the Milani mission also aims to achieve several technological objectives related to testing environmental effects on key hardware and validating novel navigation algorithms. The different phases of the Milani mission are shown in Figure \ref{fig:milani_phases}. One of the payloads of Milani is the NavCam, see Figure  \ref{fig:milani_navcam}, an optical imager which nominally provides navigational information but will also be used to perform scientific investigations and perform experiments related to autonomous navigation around asteroids \cite{Pugliatti2023}. 

The Navigation Camera of Milani is based on a Tyvak Star Tracker technology, customized for the Milani mission. Optical lenses were designed and coated specifically for Milani by Optec SpA. Tyvak ProxOps Vis Imgr WFOV Enhanced camera was selected as baseline for the mission, due to FOV constraints for the primary navigation camera, namely the requirement to keep the whole system in view at all mission ranges. This imager contains an RGB filter that allows for spectrometric measurements, similar to the LUKE camera on-board the LICIACube spacecraft \cite{Poggiali2022}, and works in a spectral range of 400–700 {[}nm{]}. Observations made by the Milani NavCam will improve the shape and rotational models, infer regolith and boulder properties, determine shocked and weathered regions on the surface, and contribute to gravity field estimates with additional navigation observables and serendipitous particle tracking data.

Some of the most important properties of the Milani and Juventas NavCam are summarized in Table \ref{tab:navcam_prop}.

\begin{table}[htbp]
\centering
\caption{Milani and Juventas NavCam Properties.}
{\fontsize{10}{12}\selectfont 
\begin{tabular}{lll}
\hline
\textbf{Parameter} & \textbf{Milani NavCam} & \textbf{Juventas NavCam} \\ 
\hline
Field of View (deg)          & 21.0 x 16.0   & 38.0 x 28.5     \\ 
\hline
Field of View (m)            & \begin{tabular}[c]{@{}l@{}} 185 @ 500 m distance \\ 740 @ 2 km distance  \\ 3700 @ 10 km distance  \end{tabular}  & \begin{tabular}[c]{@{}l@{}}
340 m @ 500 m distance \\ 1370 m @ 2 km distance \\ 2270 m @ 3.3 km distance 
\end{tabular}

\\                                

\hline
Image size (pixels)          & 2048 x 1536 & 2048 x 1536                                                 
\\ \hline
Pixel size (µm)              & 2.2 x 2.2 & 5.8 x 5.7                                                                                                               \\ \hline
Focal length (mm)            & 12.96 & 17.10 mm                                                                                                                    \\ \hline
Working distance (km)        & 30–0.2    & 30–0.2                                                                                     \\ \hline

Resolution (cm/px)           & \begin{tabular}[c]{@{}l@{}} 12 @ 500m distance \\  46 @ 2km distance \\ 232 @ 10km distance \end{tabular} & \begin{tabular}[c]{@{}l@{}} 16 @ 500m distance \\ 67 @ 2km distance \\ 110 @ 3.3km distance  
\end{tabular}  \\
\hline
\end{tabular}
}
\label{tab:navcam_prop}
\end{table}

\section{Methods}
A gravity RSE is a particular application of the OD process, which aims to estimate a set of parameters that affect the motion of the spacecraft. While the focus of the OD, as a part of the more general process of navigation, is the trajectory of the spacecraft, the focus of gravity science is also the accurate modelling and estimation of the physical parameters of interest, such as the gravity field, the rotational state, and the orbits of celestial bodies. When dealing with future missions, the use of a realistic setup and assumptions within the numerical simulations allows inferring, before the execution of the RSE, the expected formal accuracies on the estimation of scientific parameters of interest, performing a so-called covariance (sensitivity) analysis \cite{bierman1977factorization}.

The covariance analysis performed in this research adopts a multi-arc approach and is based on a linearized weighted least-squares principle, where the problem is posed in a single batch. The OD simulations are performed using NASA-JPL’s orbit determination program MONTE (Mission Analysis, Operations, and Navigation Toolkit Environment), currently used for the operations of all NASA’s space missions managed by JPL \cite{evans2018monte}, as well as for radio science data analysis and processing.

Given the matrix of measurement partials \textbf{H} with respect to the estimated parameters, and the measurement weight matrix \textbf{W}, the covariance matrix \textbf{P} at time $t_0$ (i.e. the filter solution epoch) can be evaluated, namely\cite{milani2010}:
\begin{equation}
        \mathbf{P}_{t_0} = \left[ \mathbf{H}_{t_0}^T \mathbf{W} \mathbf{H}_{t_0} + \mathbf{P}_0^{-1} \right]^{-1},	
\end{equation}
where $\mathbf{P}_0$ represents the a priori covariance matrix, which reflects the confidence in the a priori knowledge of each global and local estimated parameter. The matrix $\mathbf{P}_{t_0}$ is the covariance matrix and represents the main output of a covariance analysis as it provides the expected accuracy of all the global and local parameters estimated in the filtering process, given at the reference epoch $t_0$.\cite{gramigna2024hera}

\section{Covariance Analyses Setup}
The simulations conducted in this work follow the framework outlined by Gramigna et al. (2024) \cite{gramigna2024hera,gramigna2024heraAAS}, particularly concerning the dynamical model, measurement model, and filter setup, which are not repeated here for brevity. Regarding the measurements, the nominal scenario includes Hera-to-Earth two-way Doppler and range, Hera-to-CubeSats inter-satellite link (ISL) two-way ranging, and Hera optical images. Building on this, we expand the dataset in this study by incorporating CubeSat opnav data.

The simulation covers the nominal asteroid phase of the Hera mission, spanning the Early Characterization Phase (ECP), Detailed Characterization Phase (DCP), and Close Observation Phase (COP). In this framework, the CubeSats are released at the end of ECP, and their nominal mission is simulated. The experimental phase of Hera, Juventas, and Milani includes Hera's closer flybys of the system, low-altitude SSTO for Milani (see Figure \ref{fig:milani_phases}), and the CubeSats' landings on the asteroids. These will be explored in a subsequent dedicated study.

Specifically, two dynamical scenarios of the Didymos system are considered, both based on the Full Two-Body Problem (F2BP) formulation. The first scenario corresponds to the F2BP excited post-impact (non-chaotic) case, as described in Section 10.1 of Gramigna et al. (2024) \cite{gramigna2024hera}. The second scenario models the F2BP post-DART impact tumbling dynamics, referred to as case-1 in Pravec et al. (2024) and Gramigna et al. (2024) \cite{pravec2024rotational,gramigna2024heraAAS}. The objective is to evaluate the improvement in navigation performance and the estimation of scientific parameters enabled by the CubeSats’ opnav data, depending on the dynamical state of the Didymos system upon Hera’s arrival in late 2026.

This investigation comprises two studies. The first establishes a reference case defined by fixed values for the CubeSats optical images’ frequency, noise levels, and pointing errors. This serves to demonstrate the contribution of the CubeSats’ opnav data to the nominal scenario, which excludes these data. The second study involves a sensitivity analysis, wherein key parameters related to the CubeSats’ opnav are varied individually. These parameters include image frequency, noise levels, and pointing errors (reflecting the CubeSats' attitude uncertainties estimated for each image). This analysis aims to assess the impact of each parameter on the overall results.

The assumptions regarding the CubeSats' optical images used in the parametric and sensitivity analyses are summarized in Table \ref{tableparams}. The reference study, which highlights the contribution of the additional CubeSat data, is based on the parameter values marked with an asterisk in Table \ref{tableparams} and incorporates images from both CubeSats. The pointing errors for the CubeSats are derived from the latest Hera CubeSats operational mission analysis report, resulting in a standard value of 0.13 deg (1$\sigma$). A best-case scenario assumes Hera-like pointing errors of 0.01 deg, while a worst-case scenario assumes 0.5 deg.

\begin{table}[]
    \centering
    \caption{CubeSat opnav simulation parameters. The parameters adopted in the CubeSats opnav reference scenario are marked with an asterisk.}
    {\fontsize{10}{12}\selectfont 
    \begin{tabular}{lll}
         \hline
         \textbf{Parameters} & \textbf{Values} & \textbf{Units}  \\
         \hline
         Image frequency & 1.0, 3.0*, 12.0 & hours \\
         Noise (sample/line) & 0.1, 1.0, 2.0* & pixels \\
         Pointing errors & 0.01, 0.13*, 0.5 & deg \\
         Opnavs data & Milani only & // \\
         &             Juventas only & \\
         &             Milani+Juventas* & \\
         \hline
    \end{tabular}
    }
    \label{tableparams}
\end{table}

In the reference scenario analyzed here, we consider optical images acquired during CubeSats navigation. Juventas primarily focuses on Didymos, capturing surface landmarks while in low-altitude SSTO orbits around the primary asteroid. It also observes Dimorphos' centroid when within the NavCam's FOV. Milani’s imaging targets are determined by the primary object fully visible in its FOV at the time, including surface landmarks and, when present, the secondary target’s centroid. If the apparent diameter of the main target exceeds 100 pixels, surface features are imaged and measured; otherwise, only centroiding is performed. Due to Milani’s NavCam characteristics, the reference scenario predominantly involves imaging Didymos, including its surface landmarks and, when visible, Dimorphos' centroid. An exception occurs during Milani’s second close flyby in the COP phase (see Figure \ref{fig:milani_phases}), where Milani targets Dimorphos since Didymos appears too large in the FOV, capturing a few images of Dimorphos' surface landmarks. In the final section of this work, we examine the potential benefits of increasing Dimorphos imaging opportunities for both Milani and Juventas.

\section{Results}
This section presents the outcomes of the reference study, which demonstrates the benefits of incorporating CubeSats opnav data into the OD process. Specifically, the study evaluates the impact of using only Milani opnav data, only Juventas opnav data, and a combination of both. In the following the nominal scenario is the standard simulation without CubeSats images, while the reference scenario is the nominal + CubeSats opnavs. In particular, within this reference analysis, the main parameters for Juventas and Milani optical imaging are as follows: each CubeSat captures one image every 3 hours, with an opnav noise of 2.0 pixels (in sample and line) and pointing errors of 0.13 deg per image. The targets of the images are described at the end of the previous Section.

\begin{figure}[htbp]
    \centering
    \includegraphics[width=\textwidth]{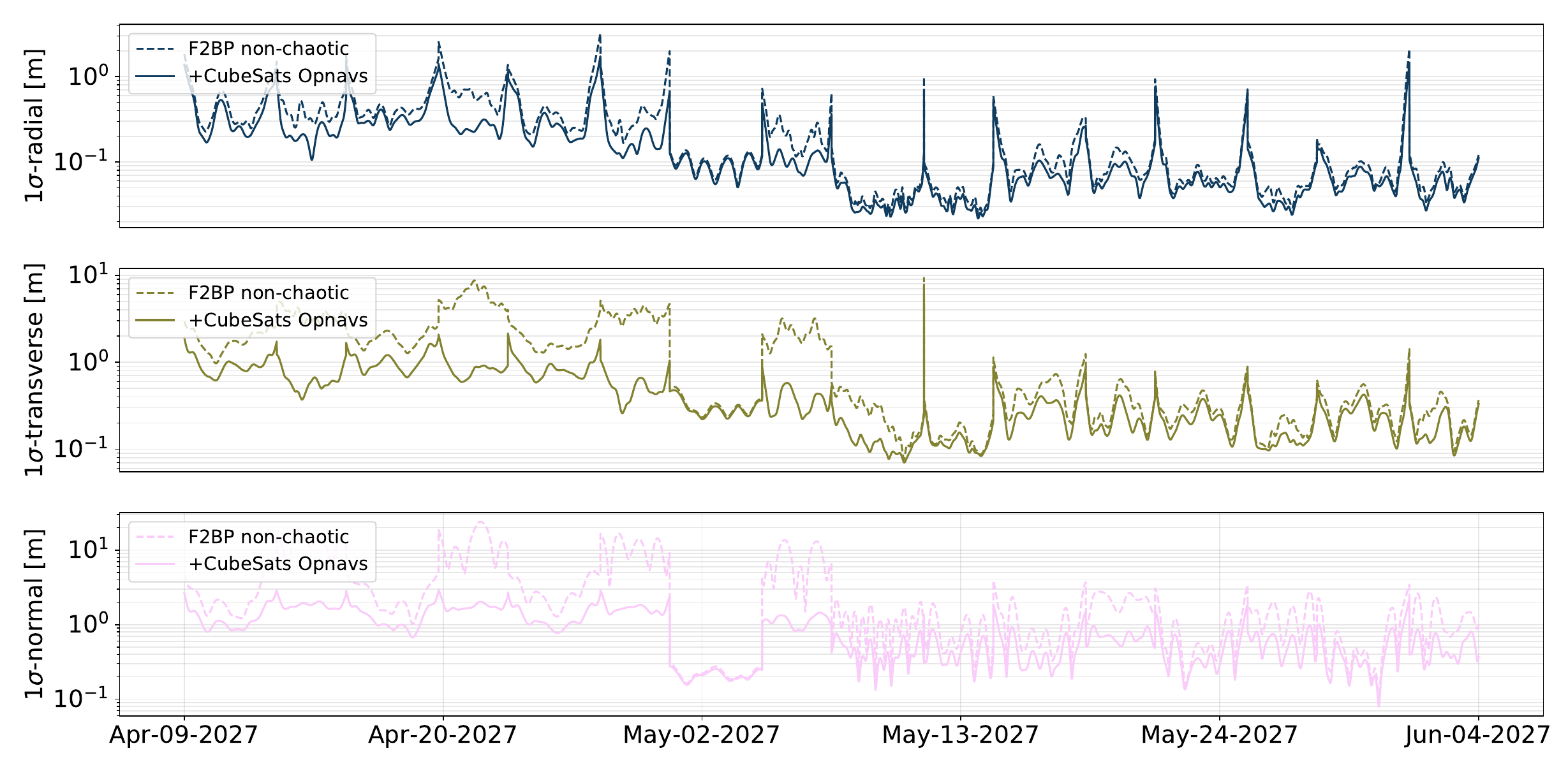}
    \caption{Expected position uncertainty (1$\sigma$) of Juventas relative to Didymos in the radial, along-track, and normal directions for the F2BP non-chaotic scenario. The dashed line represents the case without CubeSat optical navigation (opnav) data, while the solid line includes both CubeSats' opnav data. The opnav data significantly improves position accuracy, particularly in the first SSTO at 3.3 km altitude.}
    \label{figjuvsigma}
\end{figure}

\begin{figure}[htbp]
    \centering
    \includegraphics[width=\textwidth]{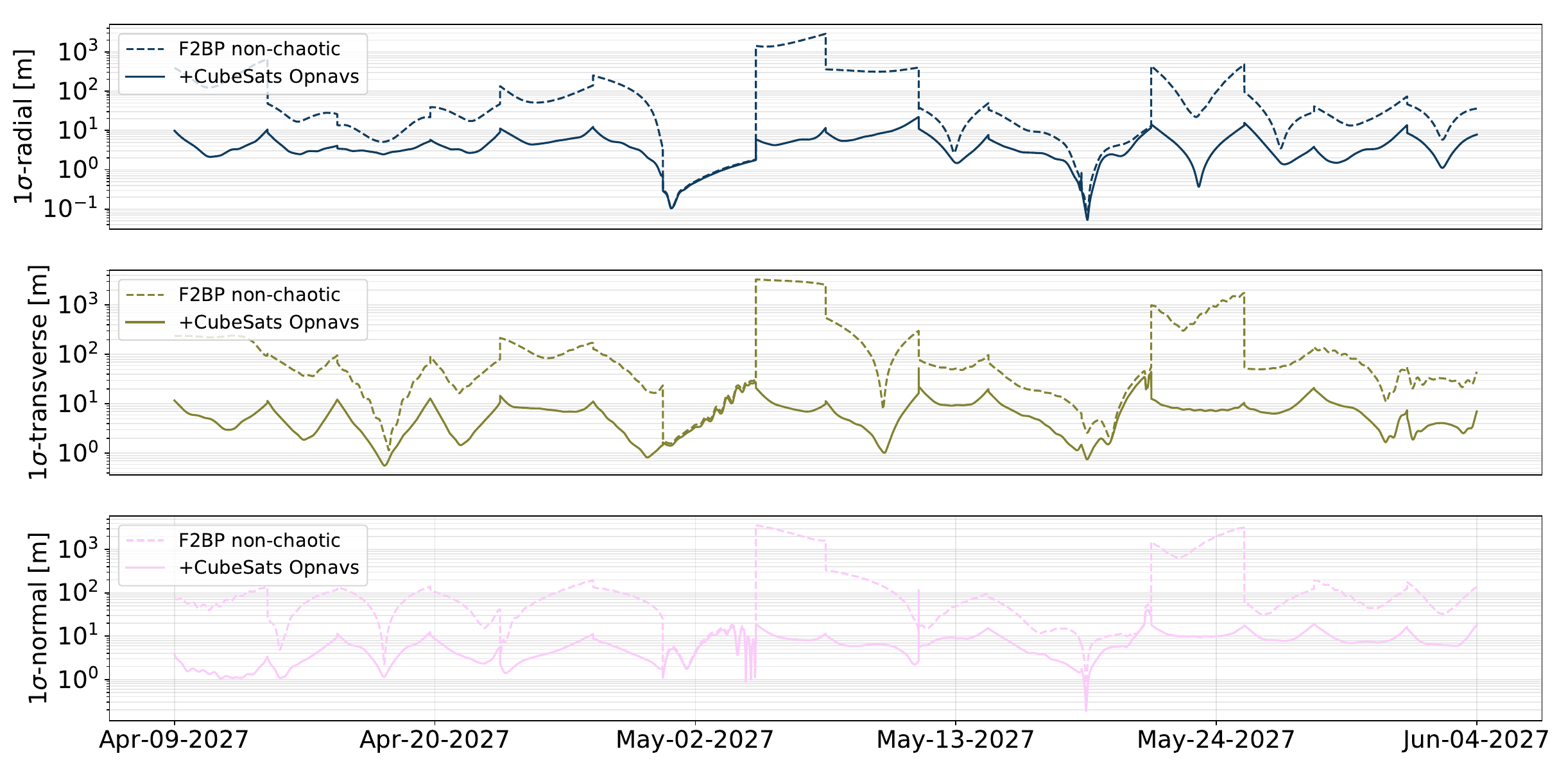}
    \caption{Expected position uncertainty (1$\sigma$) of Milani relative to Didymos in the radial, along-track, and normal directions for the F2BP non-chaotic scenario. The dashed line represents the case without CubeSat opnav data, while the solid line includes both CubeSats' opnav data. Due to Milani's higher altitudes during Hera-like hyperbolic arcs, the opnav data yields improvements of up to three orders of magnitude in all components.}
    \label{figmilsigma}
\end{figure}

Figures \ref{figjuvsigma} and \ref{figmilsigma} illustrate the reconstructed position uncertainties (1$\sigma$) of Juventas and Milani relative to Didymos in the radial, transverse, and normal (RTN) components for the excited non-chaotic dynamical scenario. The dashed lines represent the simulations without CubeSat opnav data (nominal scenario), while the solid lines include opnav data from both CubeSats (CubeSats reference scenario).

For Juventas, which operates initially in a higher SSTO at 3.3 km and later transitions to a lower SSTO at 2.0 km altitude, the opnav data significantly enhance position knowledge in the first SSTO. At this altitude, the gravity signal is weaker, making the addition of opnav data particularly beneficial, especially for the transverse and normal components, which improve by up to an order of magnitude. In the lower SSTO, the higher gravity signal from ISL ranging already improves position uncertainties. However, the inclusion of opnav data further refines the position knowledge, improving the radial and transverse components by 20–25\% and the normal component by up to 50\%. With opnav data, Juventas’ position uncertainty consistently remains at sub-meter levels throughout the mission.

For Milani, which primarily conducts Hera-like hyperbolic arcs at higher altitudes than Juventas, the impact of opnav data is even more pronounced. As shown in Figure \ref{figmilsigma}, position knowledge improves by up to three orders of magnitude across all RTN components, achieving meter and sub-meter level accuracy. This improvement is relevant for navigation, and programming with more accuracy the scientific operations for its experiments and observations.


\subsection{Enhanced Science Parameter Estimation}
The combination of increased data availability and improved knowledge of the positions of Juventas and Milani enhances the accuracy in estimating key scientific parameters, including asteroid masses, moments of inertia, rotational states, and orbital elements. Notably, this also leads to an improvement in Hera's positional accuracy due to better knowledge of the estimated parameters. Specifically, the images captured by the CubeSats enable more precise triangulation of the Didymos system, effectively breaking the scaling symmetry that is the primary source of uncertainty in Hera's radial position relative to Didymos. As a result, Hera's radial position improves by up to 80\% compared to the nominal case without CubeSat images, whereas the improvements in the other two components are negligible.

Furthermore, Tables \ref{tableF2BPnonchaotic} and \ref{tableF2BPchaotic} summarize the formal uncertainties for the gravitational mass (GM) and moments of inertia (MOI) of Didymos and Dimorphos in the two dynamical scenarios analyzed: the F2BP non-chaotic and chaotic cases. In Table \ref{tableF2BPnonchaotic}, the contribution of each CubeSat's opnav data is highlighted. In terms of GMs, Juventas, operating closer to the system during the nominal phase, provides higher information content than Milani, which remains at higher altitudes in this phase. Notably, Milani will also transition to SSTOs during its experimental phase, enabling the collection of additional scientific data independent of Juventas.

The results show that Didymos’ and Dimorphos’ GMs improve respectively by factors of 1.3–1.1 for the non-chaotic case and 1.4–1.1 for the chaotic scenario. The uncertainties in the principal MOI components improve by approximately 10\% for both bodies. These results also improve the estimation of the J$_2$ gravity coefficient, with reductions in uncertainty by a factor of $\sim$1.1-1.2 for both Didymos and Dimorphos. Additionally, Didymos’ S$_{21}$ Stokes coefficient improves by 30-45\%, while Dimorphos’ C$_{22}$ coefficient improves by a factor 1.1-1.2 for the non-chaotic and chaotic cases. In this context, the higher-order spherical harmonic coefficients do not show significant improvement from incorporating the CubeSats' images.

\begin{table}[htbp]
    \caption{Estimated formal uncertainties (1$\sigma$) in the GM and MOI for the F2BP non-chaotic scenario. Results are shown for the nominal case and when including Juventas-only, Milani-only, or both CubeSats' opnav data. Values are rounded to two decimal places, with unconstrained a priori covariances.} 
    \label{tableF2BPnonchaotic}
    \centering
    {\fontsize{10}{12}\selectfont 
    \begin{tabular}{ll|llll} 
    \hline 
	\textbf{Coefficient} & \textbf{Nominal} & \textbf{F2BP} & \textbf{+ Milani-only} &  \textbf{+ Juventas-only} & \textbf{+ Juventas-Milani} \\
    & \textbf{Value} & \textbf{non-chaotic} & \textbf{opnavs} & \textbf{opnavs} & \textbf{opnavs}
    \\
    \hline 
	\textbf{Didymos} & & & & &\\
	GM (km$^3$/s$^2$) & 3.94$\cdot10^{-8}$ & 1.45$\cdot10^{-12}$ & 1.43$\cdot10^{-12}$ & 1.14$\cdot10^{-12}$ & 1.13$\cdot10^{-12}$ \\
    \textbf{MOI} (kg $\cdot$ km$^2$) & & & & &\\
	$I_{XX}$ & 2.92$\cdot10^{10}$ & 5.19$\cdot10^{9}$ & 5.10$\cdot10^{9}$ & 5.01$\cdot10^{9}$ & 4.93$\cdot10^{9}$ \\
    
	$I_{YY}$ & 3.14$\cdot10^{10}$ & 5.19$\cdot10^{9}$ & 5.10$\cdot10^{9}$ & 5.01$\cdot10^{9}$ & 4.93$\cdot10^{9}$ \\
    
    $I_{ZZ}$ & 3.92$\cdot10^{10}$ & 5.19$\cdot10^{9}$ & 5.10$\cdot10^{9}$ & 5.01$\cdot10^{9}$ & 4.93$\cdot10^{9}$ \\
    
	$I_{XY}$ & 0.00 & 1.40$\cdot10^{7}$ & 1.40$\cdot10^{7}$ & 1.40$\cdot10^{7}$ & 1.40$\cdot10^{7}$ \\  
    
    $I_{XZ}$ & 0.00 & 4.08$\cdot10^{7}$ &  4.08$\cdot10^{7}$ & 4.08$\cdot10^{7}$ & 4.08$\cdot10^{7}$ \\
    
	$I_{YZ}$ & 0.00 & 3.36$\cdot10^{6}$ &  3.15$\cdot10^{6}$ & 1.91$\cdot10^{6}$ & 1.87$\cdot10^{6}$ \\  
    
	\hline
	\textbf{Dimorphos} & & & & &\\
	GM (km$^3$/s$^2$) & 3.45$\cdot10^{-10}$ & 3.35$\cdot10^{-13}$ & 3.31$\cdot10^{-13}$ & 3.03$\cdot10^{-13}$ & 3.01$\cdot10^{-13}$ \\
    \textbf{MOI} (kg $\cdot$ km$^2$) & & & & &\\
	$I_{XX}$ & 1.08$\cdot10^{7}$ & 5.36$\cdot10^{5}$ & 5.27$\cdot10^{5}$ & 4.91$\cdot10^{5}$ & 4.85$\cdot10^{5}$\\
    
	$I_{YY}$ & 1.12$\cdot10^{7}$ & 5.57$\cdot10^{5}$ & 5.48$\cdot10^{5}$ & 5.11$\cdot10^{5}$ & 5.05$\cdot10^{5}$\\
    
    $I_{ZZ}$ & 1.54$\cdot10^{7}$ & 7.64$\cdot10^{5}$ & 7.52$\cdot10^{5}$ & 7.01$\cdot10^{5}$ & 6.92$\cdot10^{5}$\\
    
	$I_{XY}$ & 0.00 & 3.12$\cdot10^{3}$ &  3.11$\cdot10^{3}$ & 3.12$\cdot10^{3}$ & 3.11$\cdot10^{3}$\\
    
    $I_{XZ}$ & 0.00 & 2.70$\cdot10^{4}$ &  2.70$\cdot10^{4}$ & 2.70$\cdot10^{4}$ & 2.70$\cdot10^{4}$ \\
    
	$I_{YZ}$ & 0.00 & 1.12$\cdot10^{4}$ &  1.11$\cdot10^{4}$ & 1.12$\cdot10^{4}$ & 1.11$\cdot10^{4}$ \\
    
    \hline
   \end{tabular}
   }
\end{table}

\begin{table}[htbp]
    \caption{Estimated formal uncertainties (1$\sigma$) in the GM and MOI for the F2BP chaotic scenario. Improvements provided by the addition of both CubeSats' opnav data are highlighted in the last column. Values are rounded to two decimal places, with unconstrained a priori covariances.} 
    \label{tableF2BPchaotic}
    \centering
    {\fontsize{10}{12}\selectfont 
    \begin{tabular}{ll|ll} 
    \hline 
	\textbf{Coefficient} & \textbf{Nominal} &  \textbf{F2BP} & \textbf{+ CubeSats} \\
    & \textbf{Value} & \textbf{chaotic} & \textbf{opnavs}
    \\
    \hline 
	\textbf{Didymos} & & &\\
	GM (km$^3$/s$^2$) & 3.94$\cdot10^{-8}$ &  2.15$\cdot10^{-12}$ & 1.50$\cdot10^{-12}$ \\
    \textbf{MOI} (kg $\cdot$ km$^2$) & & & \\
	$I_{XX}$ & 2.92$\cdot10^{10}$ &  2.31$\cdot10^{8}$ & 2.21$\cdot10^{8}$\\
	$I_{YY}$ & 3.14$\cdot10^{10}$ &  2.31$\cdot10^{8}$ & 2.21$\cdot10^{8}$\\
    $I_{ZZ}$ & 3.92$\cdot10^{10}$ &  2.31$\cdot10^{8}$ & 2.21$\cdot10^{8}$\\
	$I_{XY}$ & 0.00 & 3.97$\cdot10^{6}$ & 3.96$\cdot10^{6}$\\  
    $I_{XZ}$ & 0.00 & 9.26$\cdot10^{6}$ & 9.26$\cdot10^{6}$\\
	$I_{YZ}$ & 0.00 & 3.72$\cdot10^{6}$ & 2.20$\cdot10^{6}$\\    
	\hline
	\textbf{Dimorphos} & & & \\
	GM (km$^3$/s$^2$) & 3.45$\cdot10^{-10}$ &  5.10$\cdot10^{-13}$ & 4.46$\cdot10^{-13}$ \\
    \textbf{MOI} (kg $\cdot$ km$^2$) & & & \\
	$I_{XX}$ & 1.08$\cdot10^{7}$ &  2.81$\cdot10^{5}$ & 2.27$\cdot10^{5}$\\
	$I_{YY}$ & 1.12$\cdot10^{7}$ &  3.76$\cdot10^{5}$ & 3.03$\cdot10^{5}$\\
    $I_{ZZ}$ & 1.54$\cdot10^{7}$ &  4.33$\cdot10^{5}$ & 3.49$\cdot10^{5}$\\
	$I_{XY}$ & 0.00 & 2.08$\cdot10^{4}$ & 2.08$\cdot10^{4}$\\  
    $I_{XZ}$ & 0.00 & 3.05$\cdot10^{4}$ & 3.05$\cdot10^{4}$\\
	$I_{YZ}$ & 0.00 & 4.63$\cdot10^{3}$ & 4.54$\cdot10^{3}$\\
    \hline
   \end{tabular}
   }
\end{table}

\subsection{Asteroids Dynamical State Improvements}

\begin{figure}[htbp]
    \centering
    \begin{subfigure}
        \centering
        \includegraphics[width=0.7\textwidth]{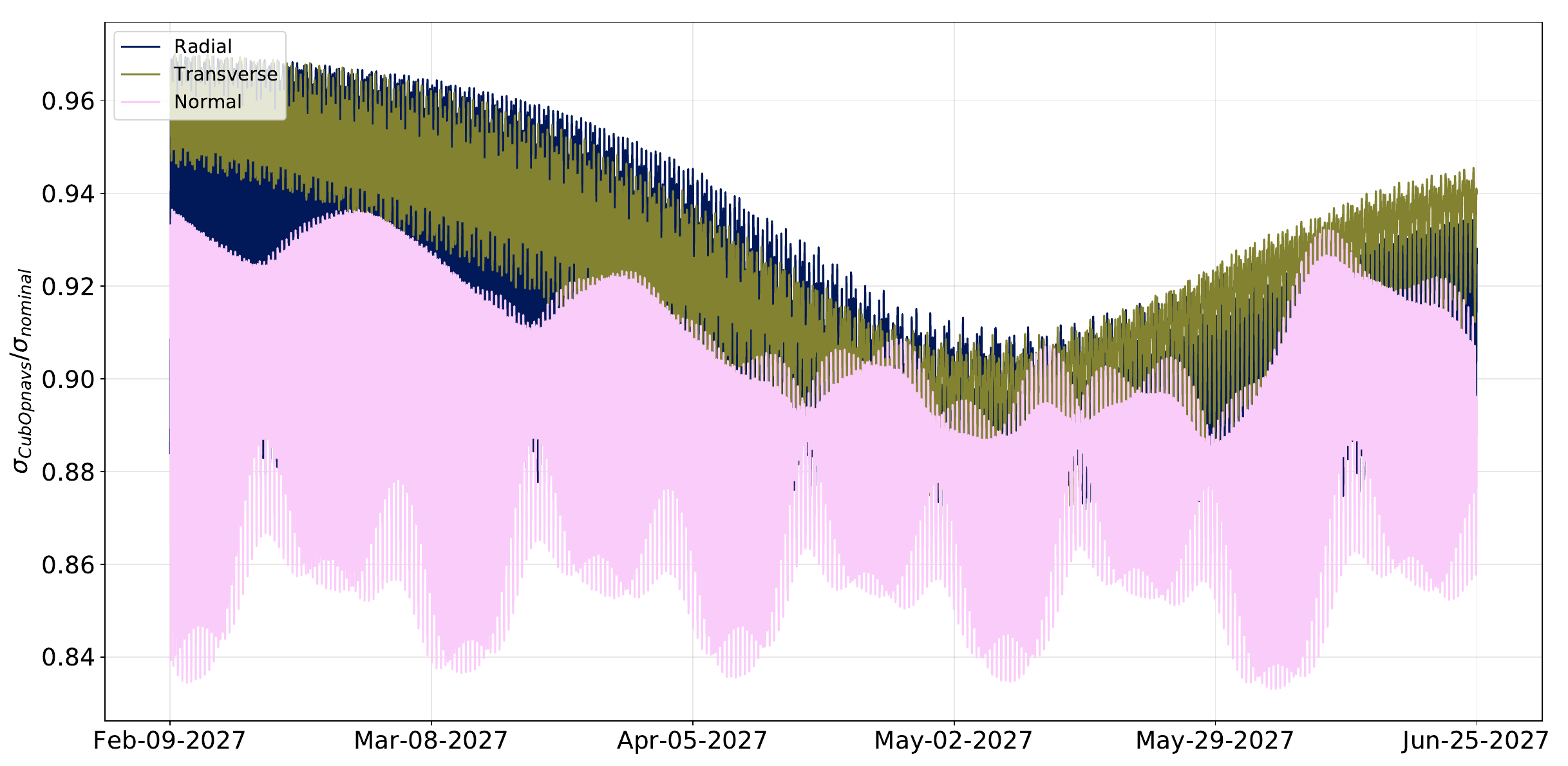} 
    \end{subfigure}
    
    \vspace{5mm} 
    
    \begin{subfigure}
        \centering
        \includegraphics[width=0.7\textwidth]{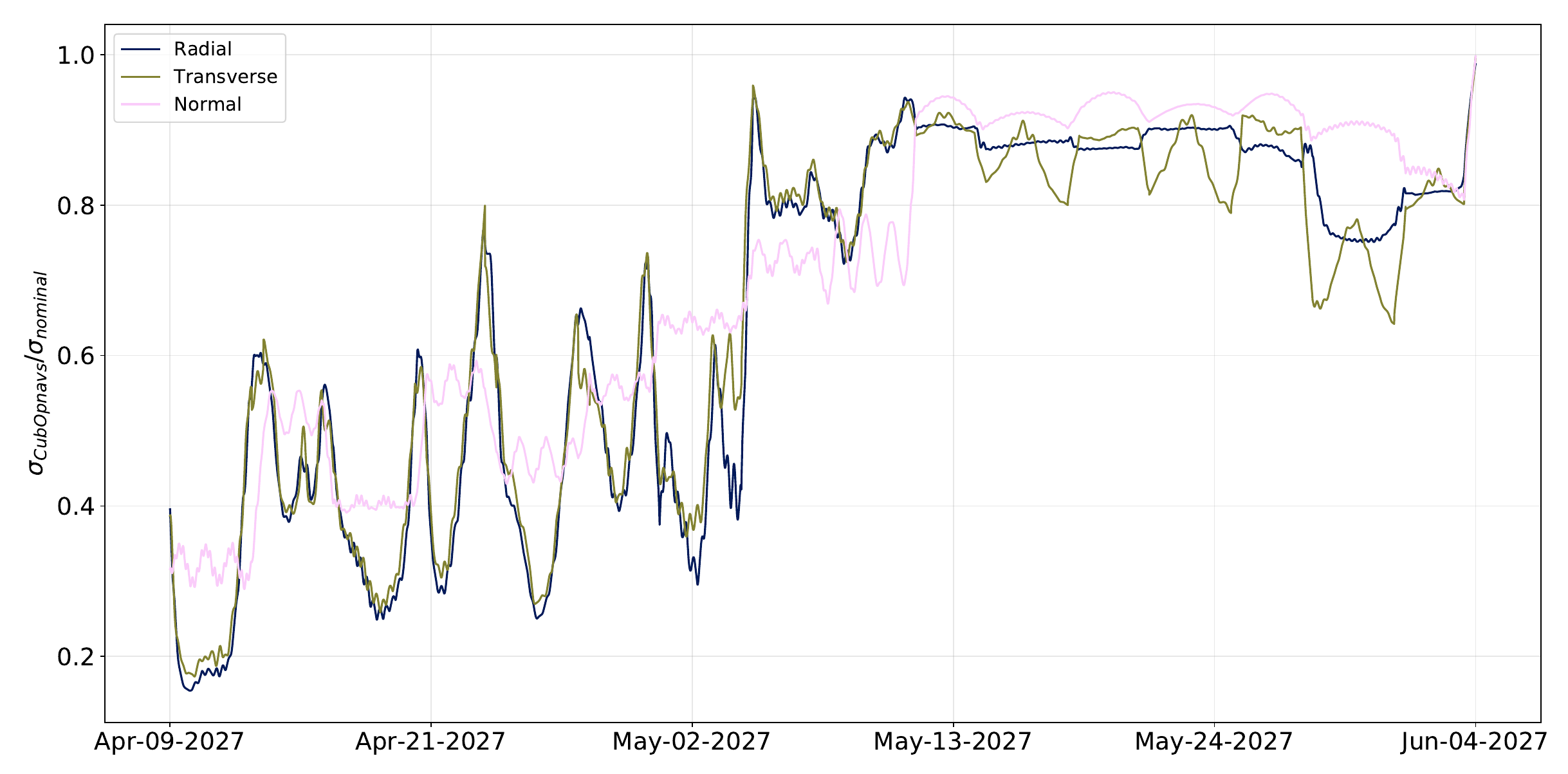} 
    \end{subfigure}       
    \caption{Relative improvement in Dimorphos' orbit accuracy due to CubeSat opnav data for (top panel) the F2BP non-chaotic scenario and (bottom panel) the F2BP chaotic scenario. The improvement is most significant during the first month after CubeSat release, reaching up to 80\% for the chaotic case and 15\% for the non-chaotic case in the normal direction.}
    \label{figdimoimprovement}
\end{figure}

Figure \ref{figdimoimprovement} illustrates the improvement in Dimorphos’ orbit knowledge achieved by incorporating CubeSat opnav data in both the non-chaotic (top panel) and chaotic (bottom panel) scenarios. This improvement is primarily attributed to the Dimorphos centroid present in the opnav data from both Juventas and Milani, with the exception of a few dedicated Dimorphos images collected by Milani.

For the non-chaotic case, where a global orbit estimation is feasible for the entire mission, the addition of CubeSat images improves Dimorphos’ position knowledge by an average of 10\%, with peak improvements of up to 15\% in the normal component. The absolute orbit knowledge for Dimorphos is refined to a precision of 10–20 cm throughout the nominal asteroid phase.

For the chaotic case, which employs a local orbit estimation approach, CubeSat opnav data significantly reduce uncertainties. In the nominal chaotic scenario, there are strong correlations, particularly between the components of Dimorphos' integrated state and between Dimorphos' integrated state and the frame parameters. The CubeSats' imaging data effectively reduce these correlations. The correlation matrices for the nominal and CubeSat opnav scenarios are provided in Figure \ref{figscorrmatrixes} in the Appendix. Position knowledge improves by up to 80\% in the RTN components during the first month following the release of the CubeSats and shows an average improvement of 50\% over the mission duration. During the second month, as Juventas transitions to its lower SSTO and Hera conducts closer flybys, the improvement from CubeSat images becomes less pronounced. Nevertheless, it still contributes to a 10–20\% enhancement in positional accuracy.

It is worth noting that the local estimation approach in this study follows the methodology of Gramigna et al. (2024) \cite{gramigna2024heraAAS}, with unconstrained a priori covariances on position and rotational state at the start of each radio science arc. A more realistic constrained multi-arc orbit determination approach is planned to achieve a coherent estimation of Dimorphos’ orbit and rotational dynamics, further reducing the uncertainties outlined here.

As mentioned before, the imaging data from Milani and Juventas is crucial for improving the determination of the rotational state of Didymos and Dimorphos, too. Simulations show that this data is particularly relevant for the tumbling dynamical case, whereas the improvement is negligible for the frame parameters of the excited non-chaotic case, except for Didymos RA-DEC which shows improvement up to a factor 2. In the latter scenario, the more stable dynamical state of the asteroid can already be well recovered using Hera-only images. In contrast, the chaotic scenario, characterized by short time-scale variations and high-amplitude librations in yaw, pitch, and roll for Dimorphos, is more challenging to resolve. In this context, the CubeSats' images are essential. 

\begin{figure}[htbp]
    \centering
    \includegraphics[width=0.7\textwidth]{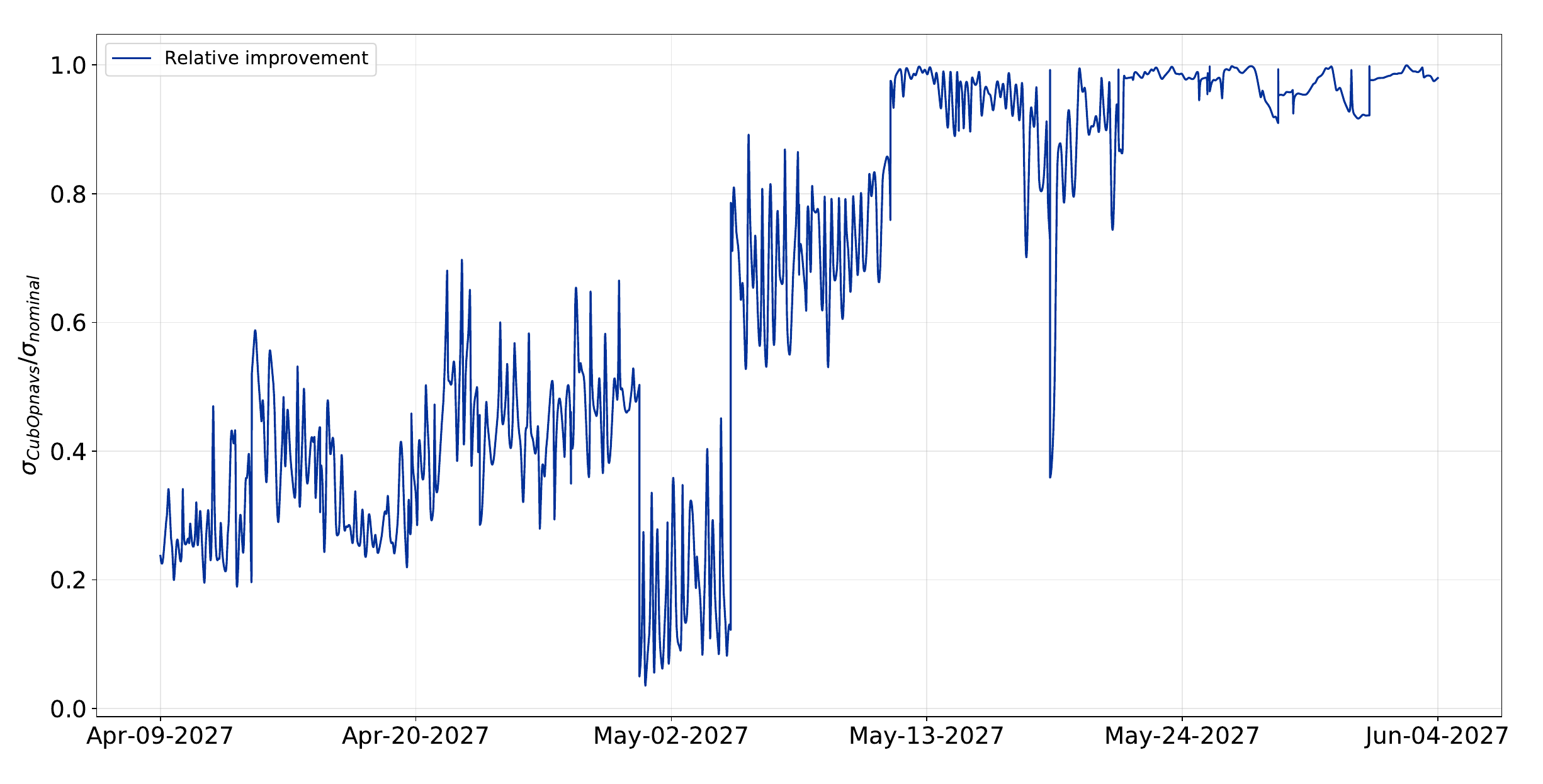}
    \caption{Relative improvement in the Z component of Dimorphos' angular velocity, expressed as $\sigma_{\text{CubeSats-opnavs}} / \sigma_{\text{nominal}}$.}
    \label{figdimoframeimprovement}
\end{figure}

Figure \ref{figdimoframeimprovement} illustrates the relative improvement provided by these measurements for the Z-component of Dimorphos' angular velocity (aligned with its pole). As shown, the relative improvement compared to the nominal case is up to 90\%, particularly during the first month when Hera performs high-altitude flybys. Typically, the rotational state is better resolved through landmark imaging. However, in this case, the improvement is primarily attributed to the inclusion of the Dimorphos centroid in the CubeSats' opnav data, as Dimorphos landmarks are imaged in only four Milani images. This centroid data, through an indirect effect, contributes to improved orbit determination and reduced correlations, as previously mentioned, ultimately leading to better accuracy in the rotational state. The improvement becomes less pronounced as Hera reduces its distance to the binary system, as previously explained. Similar improvements, both in terms of magnitude and trend, are observed for all rotational state parameters of Didymos and Dimorphos (right ascension, declination, prime meridian, and angular velocities X, Y, Z). These are not included here to avoid redundancy, as the plots would appear similar.

\begin{figure}[htbp]
    \centering
    \includegraphics[width=0.8\textwidth]{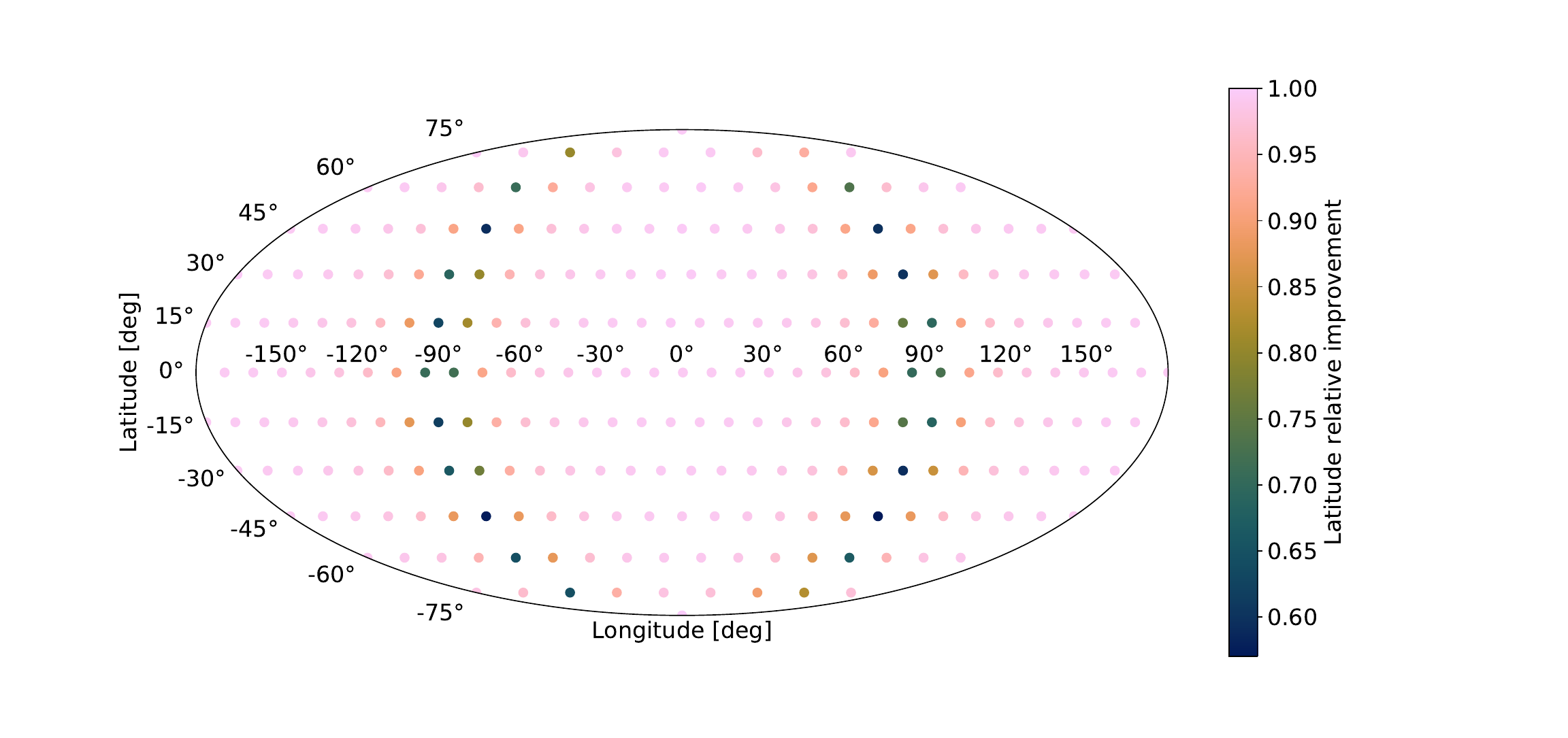}
    \caption{Relative improvement in the latitude coordinate of Didymos landmarks, expressed as $\sigma_{\text{Lat-CubeSats-opnavs}} / \sigma_{\text{Lat-nominal}}$.}
    \label{figdidylandmarksrelimp}
\end{figure}

For Dimorphos landmarks, the few Milani images results in an improvement of less than 5\% across all radius, longitude, and latitude coordinates. The resulting average 1$\sigma$ formal uncertainties (across 200 landmarks) are: 0.77 m in radius, 0.49 deg in latitude, and 0.70 deg in longitude. In the case of Didymos landmarks, the improvement is particularly significant for the latitude coordinate. Figure \ref{figdidylandmarksrelimp} presents the relative improvement in latitude coordinate formal uncertainties, expressed as the ratio of $\sigma$ values, compared to the nominal case without CubeSats images. Notably, the improvement is concentrated in two bands, corresponding to landmarks that cannot be well-constrained by Hera images alone. Improvements of up to 40\% are observed, thanks to Juventas and Milani images. The improvement in longitude (not shown here for brevity) is about 10\%, mainly near Didymos' south pole, while the improvement in the radius parameter is below 1\%. The average formal uncertainties for Didymos landmarks, including CubeSats opnavs, are: 2.25 m in radius, 0.15 deg in latitude, and 0.42 deg in longitude.

To summarize, for the excited non-chaotic and chaotic scenarios, the CubeSats' images provide comparable improvements in CubeSats' position knowledge, asteroid masses, moments of inertia, and landmark positions. However, in the chaotic scenario, the CubeSats' optical data is particularly critical for accurately determining Dimorphos' dynamical and rotational state, which is more challenging to constrain using Hera-only images.

\subsection{Sensitivity analyses}
The sensitivity analysis was performed by varying the image frequency, CubeSats’ pointing error, and image noise to assess how these parameters influence the formal uncertainties. Specifically, the analysis involved varying one parameter at a time while keeping the others constant. The adopted values are those defined in Table \ref{tableparams}. It is important to note that the reference scenario used for comparison in the following discussion corresponds to the standard case outlined in the Results section. This scenario is characterized by a 0.13 deg pointing error, 2.0-pixel noise, and an image frequency of one image every 3 hours.

\begin{figure}[htbp]
    \centering
    \begin{subfigure}
        \centering
        \includegraphics[width=0.5\textwidth]{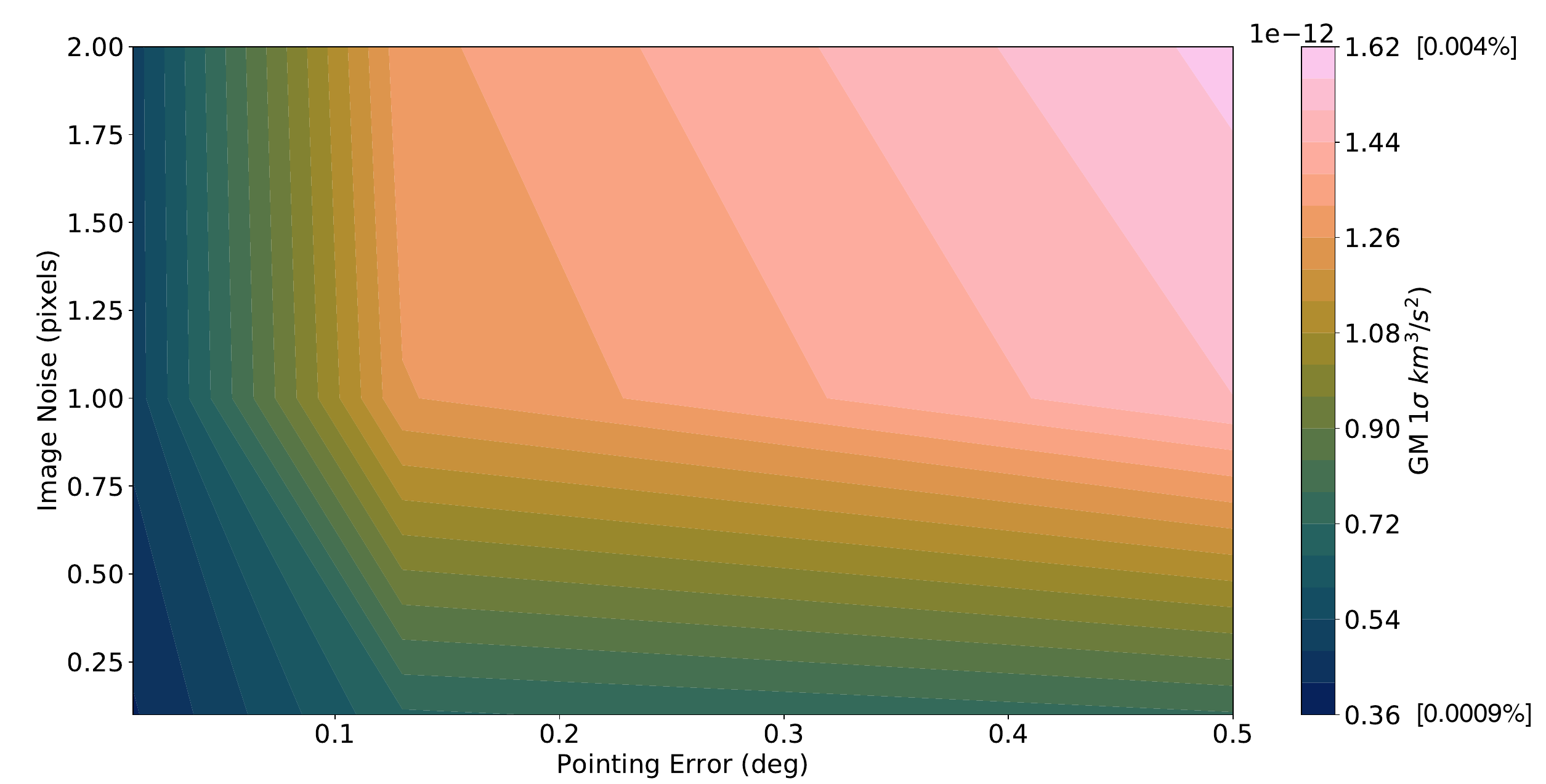} 
    \end{subfigure}
    
    \vspace{2mm} 
    
    \begin{subfigure}
        \centering
        \includegraphics[width=0.5\textwidth]{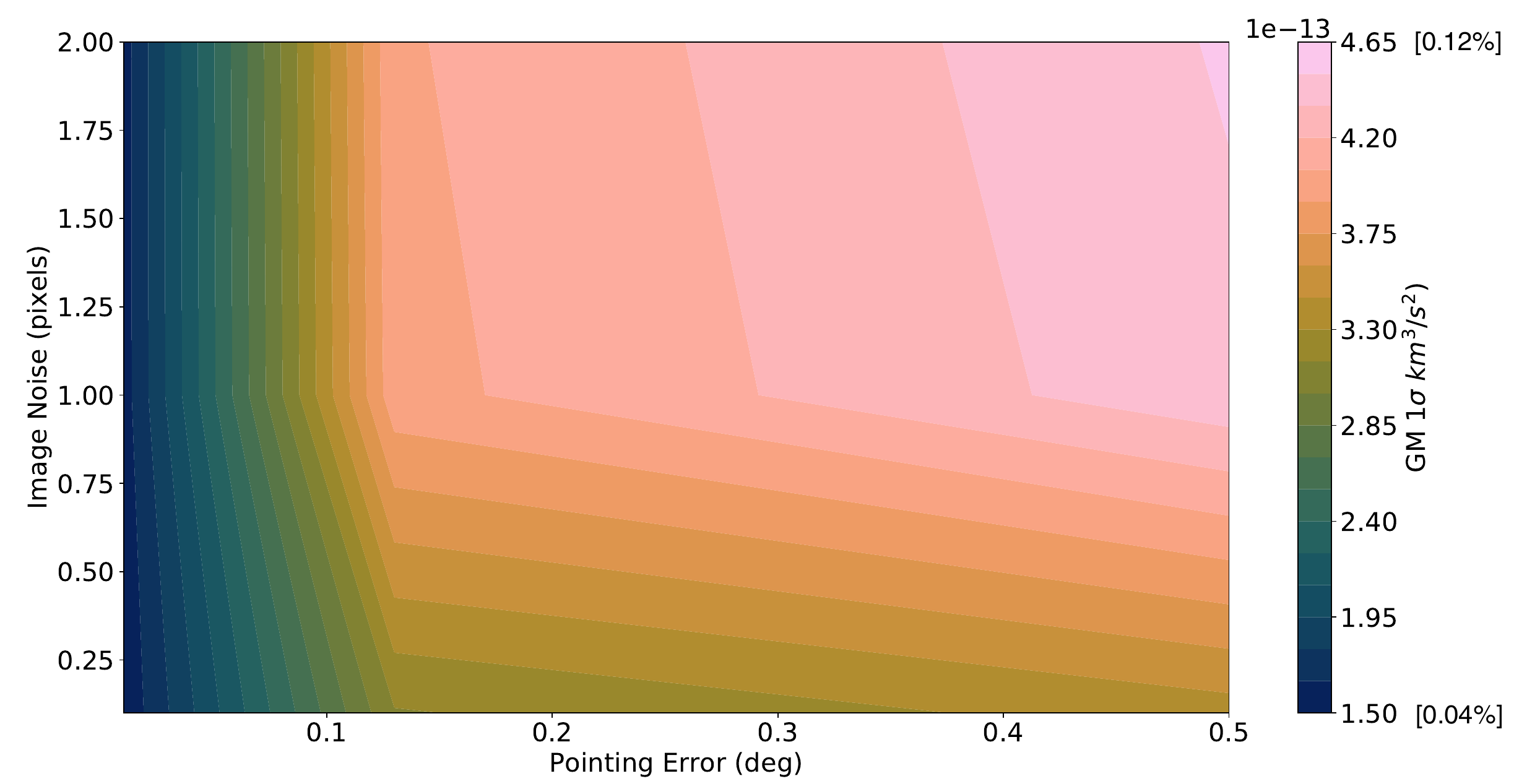} 
    \end{subfigure}   
    \caption{Formal uncertainties (1$\sigma$) in Didymos (upper panel) and Dimorphos (lower panel) GMs as a function of CubeSats pointing error (x-axis) and image noise (y-axis).}
    \label{figsGMsenspointvsnoise}
\end{figure}

First, we fixed the image frequency at 3 hours and varied the pointing error and image noise. Figure \ref{figsGMsenspointvsnoise} presents the GM uncertainty (1$\sigma$) as a contour plot for Didymos (upper panel) and Dimorphos (lower panel) as a function of pointing error and image noise. The best-case scenario, characterized by a 0.01 deg pointing error and 0.1-pixel noise (with the image frequency fixed at 3 hours), shows relative uncertainties of 0.0009\% for Didymos and 0.04\% for Dimorphos. This represents a factor of 3.3 and 2.3 improvement, respectively, compared to the CubeSats opnavs case reported in Table \ref{tableF2BPnonchaotic}. Conversely, the worst-case scenario, with a 0.5 deg pointing error and 2.0-pixel noise, results in relative uncertainties of 0.004\% for Didymos and 0.12\% for Dimorphos.

\begin{figure}[htbp]
    \centering
    \begin{subfigure}
        \centering
        \includegraphics[width=0.5\textwidth]{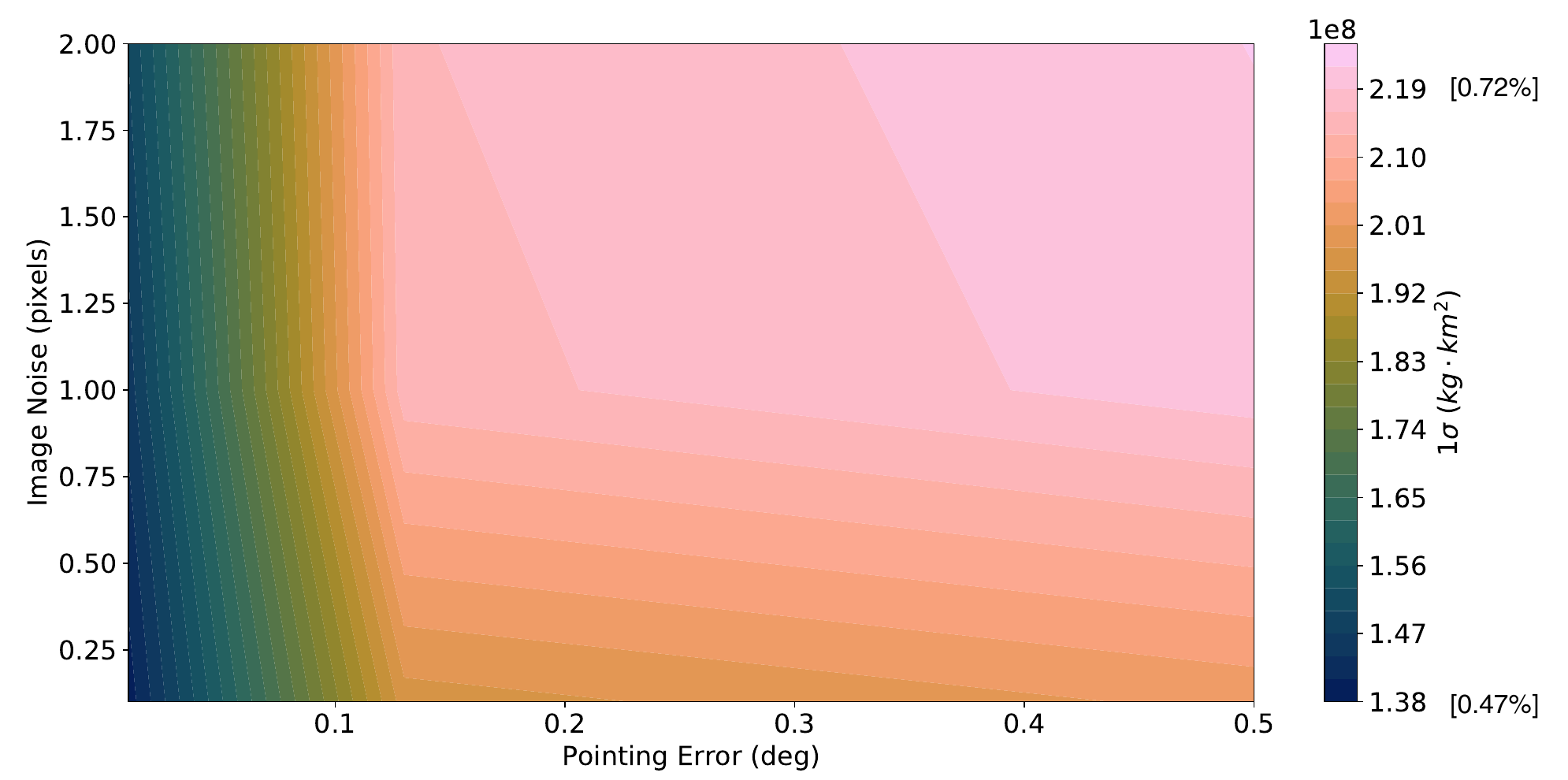} 
    \end{subfigure}
    
    \vspace{2mm} 
    
    \begin{subfigure}
        \centering
        \includegraphics[width=0.5\textwidth]{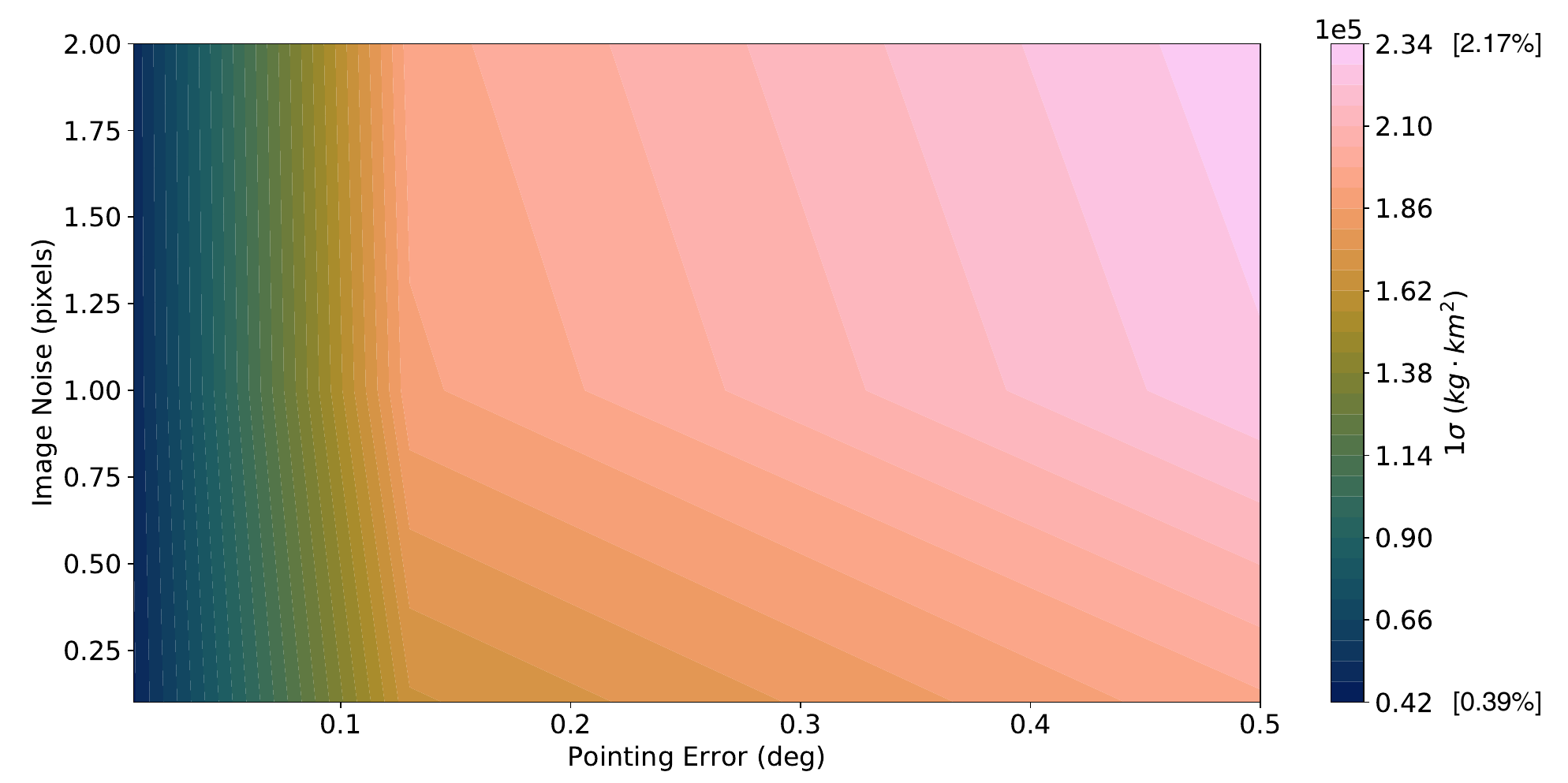} 
    \end{subfigure}   
    \caption{Formal uncertainties (1$\sigma$) in Didymos (upper panel) and Dimorphos (lower panel) I$_{XX}$ MOI as a function of CubeSats pointing error (x-axis) and image noise (y-axis).}
    \label{figsMOIsenspointvsnoise}
\end{figure}

Similarly to Figure \ref{figsGMsenspointvsnoise}, Figure \ref{figsMOIsenspointvsnoise} illustrates the formal uncertainties of the I$_{XX}$ MOI for Didymos and Dimorphos. The I$_{YY}$ and I$_{ZZ}$ components are not included for brevity, but they follow a similar trend in both the contour plots and the relative variations in formal uncertainties. For the best-case scenario, Didymos and Dimorphos show improvements of about a factor of 2 and 6, respectively, compared to the reference case (0.13 deg pointing error and 2.0-pixel noise). In contrast, the worst-case scenario (0.5 deg pointing error and 2.0-pixel noise) does not significantly degrade performance compared to the reference case.

From Figures \ref{figsGMsenspointvsnoise} and \ref{figsMOIsenspointvsnoise}, it can be observed that the primary driver of the expected accuracy in parameter estimation is the pointing error of Juventas and Milani. Proper estimation and characterization of this error will be critical. On the other hand, image noise has a less pronounced impact on the estimation of GMs and MOIs.

\begin{figure}[htbp]
    \centering
    \begin{subfigure}
        \centering
        \includegraphics[width=0.5\textwidth]{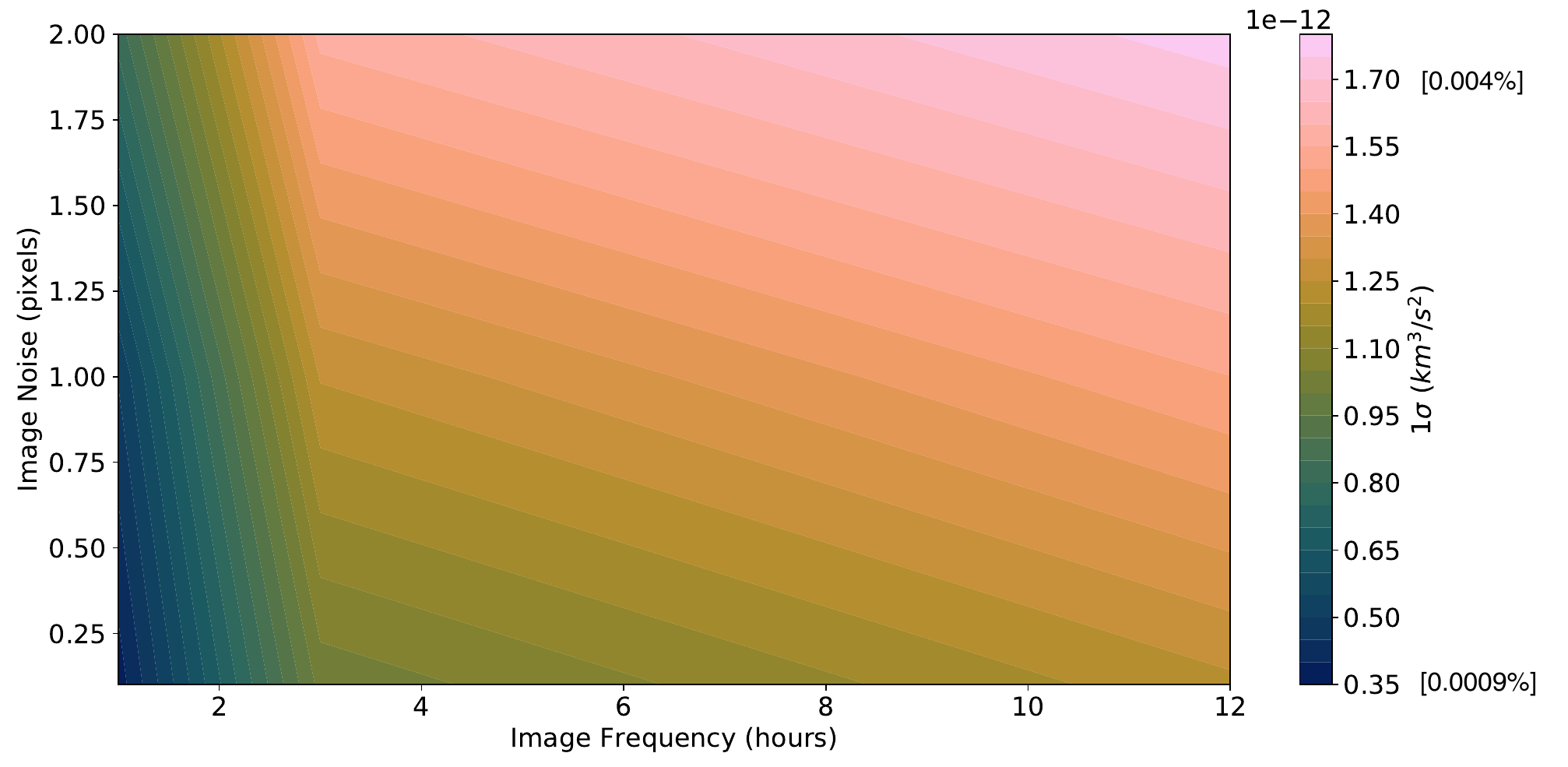} 
    \end{subfigure}
    
    \vspace{2mm} 
    
    \begin{subfigure}
        \centering
        \includegraphics[width=0.5\textwidth]{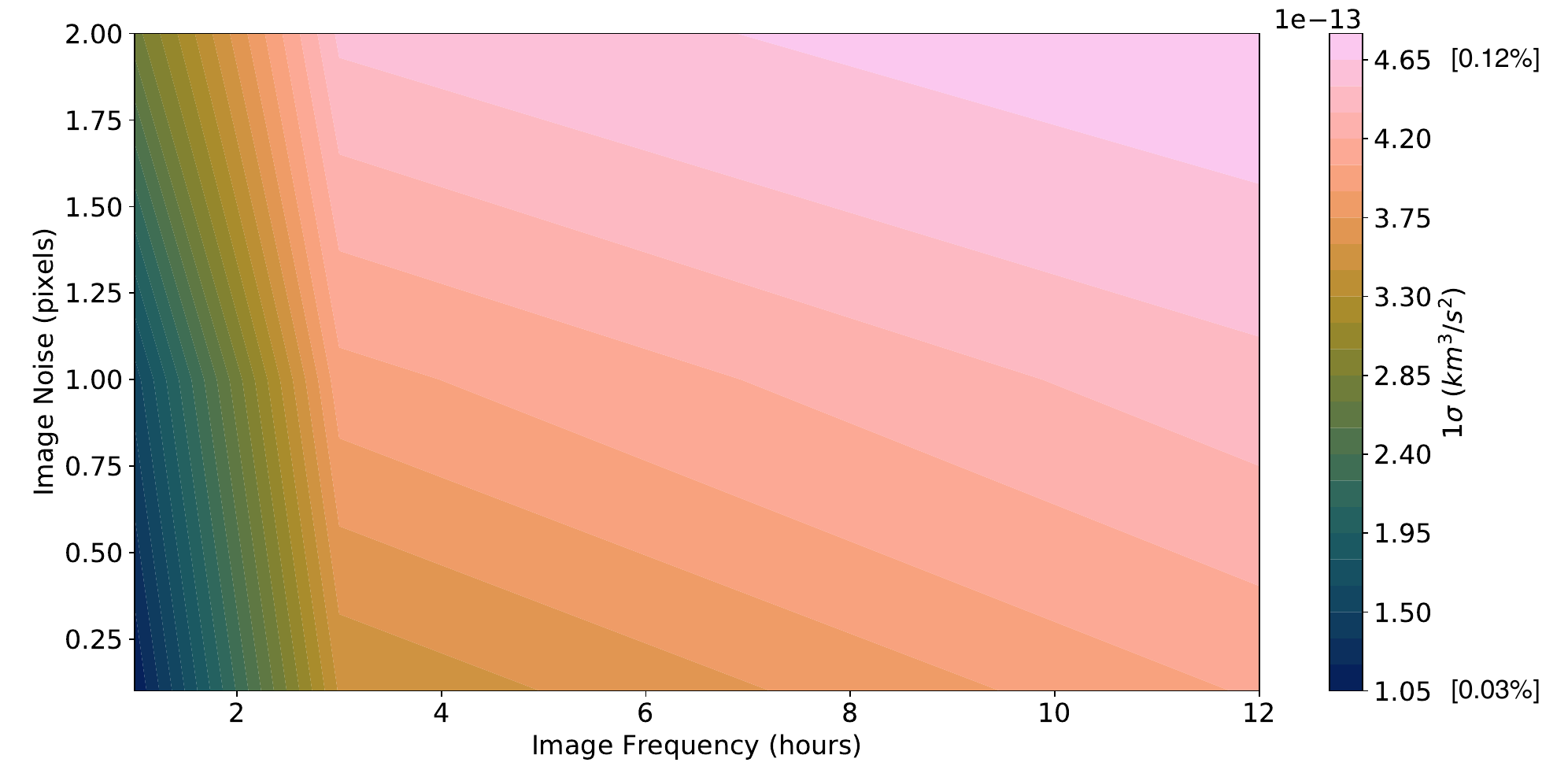} 
    \end{subfigure}   
    \caption{Formal uncertainties (1$\sigma$) in Didymos (upper panel) and Dimorphos (lower panel) GMs as a function of image frequency (x-axis) and image noise (y-axis).}
    \label{figsGMsensfreqvsnoise}
\end{figure}

Next, we kept the pointing error fixed at the reference value of 0.13 deg and varied the image frequency and image noise. The resulting contours displayed in Figure \ref{figsGMsensfreqvsnoise} exhibit a similar pattern to the previous Figures, with a factor of approximately 4 difference between the best and worst-case scenarios for both Didymos and Dimorphos GMs.

Finally it is worth mentioning the performances that the best-case studied scenario within the sensitivity analysis (0.01 deg pointing error, 1 image every hour for both CubeSats, 0.1 pixel image noise) could achieve. For this case the GMs improve by a factor 6.8 and 5.0 for Didymos and Dimorphos, respectively, with respect to the nominal scenario without CubeSats images. In addition, the principal MOI are better recovered by a factor 4.2 for Didymos and 11.1 for Dimorphos (relative uncertainties of 0.2\% for both asteroids). In this context, the overall degree-3 gravity of Didymos improve by a factor 1.3 again with respect to the reference scenario with CubeSats opnavs.

For navigation purposes this is especially relevant for the CubeSats too. In particular within this best case both Milani and Juventas further improve by one order of magnitude their orbital uncertainties if compared to the reference simulation already with CubeSats opnavs, see solid line of Figure \ref{figmilsigma}. This allows to reach formal uncertainties of sub-m level in all components throughout all the nominal asteroid phase for Milani and below the 10 cm level in all components for Juventas. This is particularly important for Milani, since it is the one characterized by higher uncertainties. 


\subsection{Milani and Juventas scientific investigation}
\label{juvscience}
The final analysis focuses on a scientific investigation conducted to optimize the planning of Juventas and Milani's scientific imaging activities. These additional Dimorphos' images could enhance our ability to refine Dimorphos' dynamical parameters and contribute to scientific studies, such as crater and surface characterization. 

While the primary imaging targets for Juventas are the surface landmarks of Didymos, its proximity to the binary system presents an opportunity to capture scientifically valuable images of Dimorphos. Similarly, Milani, which in the reference scenario was the only spacecraft to capture a limited number of Dimorphos images, could also significantly increase its scientific imaging of Dimorphos.

In these simulations, in addition to the CubeSat images described in the reference scenario (see previous section), it is assumed that Juventas will slew its NavCam to point at Dimorphos, enabling the acquisition of one image of Dimorphos, along with its surface landmarks, every five hours. Separately, for Milani, a different simulation assumes that it will capture one image of Dimorphos every three hours, but only during its two close flybys, specifically when Milani is within 5 km of Didymos. This scientific investigation was performed independently for Juventas and Milani to evaluate their individual contributions and to analyze the unique Dimorphos images each spacecraft could acquire. Simulations were performed for both the excited non-chaotic and chaotic dynamical scenarios.

\begin{figure}[htbp]
    \centering
    \begin{subfigure}
        \centering
        \includegraphics[width=0.5\textwidth]{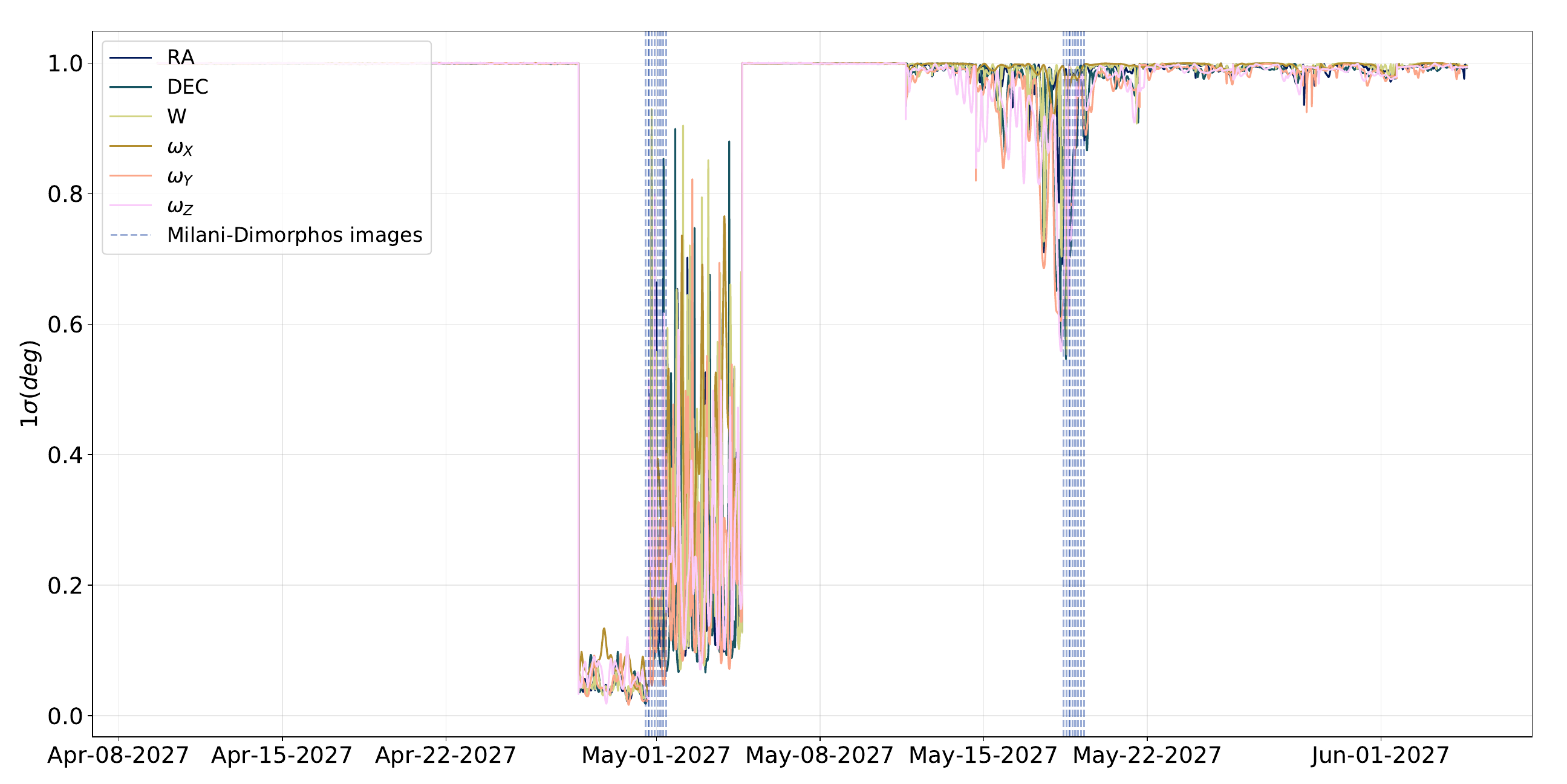} 
    \end{subfigure}
    
    \vspace{2mm} 
    
    \begin{subfigure}
        \centering
        \includegraphics[width=0.5\textwidth]{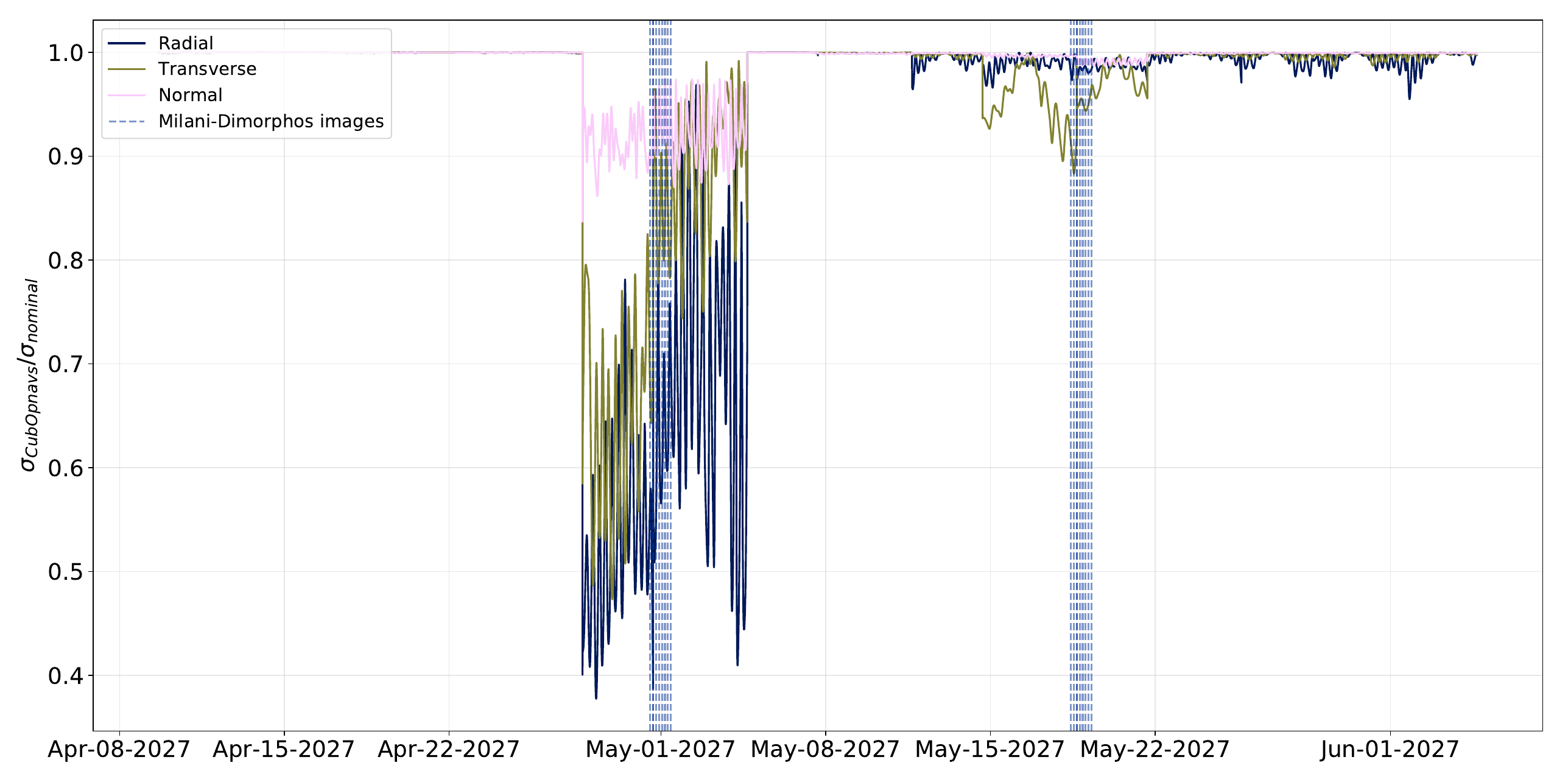} 
    \end{subfigure}   
    \caption{Relative improvement from Milani-Dimorphos images captured during the two close flybys, evaluated against the CubeSats' opnav reference scenario. Top panel: Improvement in Dimorphos rotational state parameters. Bottom panel: Improvement in Dimorphos' orbit, expressed in RTN components relative to Didymos.}
    \label{figmilscientinv}
\end{figure}

\textit{Milani}\\
Figure \ref{figmilscientinv} illustrates the additional improvement in Dimorphos's dynamical state provided by the Milani-Dimorphos images during close flybys, particularly for the chaotic case, compared to the CubeSats' opnav reference scenario. Notably, the enhancement in frame parameters is significant, reaching up to 100\% in the DCP when Hera is still relatively far from the system. This improvement is most pronounced near the time the images are obtained, as the local estimation approach primarily enhances parameters close to the acquisition times.

The second set of images also contributes to improvements, albeit less effectively, due to the already highly accurate images captured during Hera's close flybys. Similarly, Dimorphos's orbit exhibits a comparable trend, with accuracy gains of up to 50\% in the along-track and radial components.

For the non-chaotic case, the improvements are less pronounced compared to the chaotic case. However, even a few images from Milani result in a 20\% enhancement in Dimorphos's $\omega_Z$ throughout the mission. This is due to the global estimation approach, which backpropagates the improvement to earlier times. In contrast, other frame parameters and orbital components show less than 5\% improvement.

\textit{Juventas}\\
In the non-chaotic scenario, adding Juventas-Dimorphos images results in slight, though not dramatic, further improvements in the radial and transverse components, with gains of up to 8\% compared to the CubeSat-navigation case shown in Figure \ref{figdimoimprovement}. However, Juventas' images significantly enhance the characterization of Dimorphos' rotational state. Improvements of up to 15\% are observed for RA-DEC-W and up to 50\% for the angular velocities in the Y-Z components, which is particularly valuable for understanding Dimorphos' excited state and potential librations. Within this scenario, the average formal uncertainties in rotational parameters are as follows: 0.83 deg in RA, 0.24 deg in DEC, and 0.88 deg in W; and angular velocities of $5.1 \cdot 10^{-5}$ deg/s in $\omega_X$, $1.9 \cdot 10^{-5}$ deg/s in $\omega_Y$, and $1.7 \cdot 10^{-6}$ deg/s in $\omega_Z$. Improvements in the GMs and principal MOI are negligible for this case.

In contrast, as observed for Milani, the chaotic dynamical scenario shows more pronounced improvements. For this case, the GMs of Didymos and Dimorphos improve by factors of 1.2 and 1.1, respectively. Similarly, the principal MOIs improve by factors of 1.2 for Didymos and 1.8 for Dimorphos. The degree-2 gravity parameters of Dimorphos also show significant enhancements, with accuracies improving by factors of 1.8 for J$_2$, 1.3 for C$_{22}$, and 1.3 for S$_{21}$.

Regarding Dimorphos' dynamical state, Juventas' images improve orbital knowledge by up to 50\% during the DCP phase and up to 10\% during the COP phase, enabling sub-meter accuracy in all orbital components even in the chaotic case. Rotational parameters (RA-DEC-W and angular velocities) also benefit significantly, with up to 50\% improvement during DCP and approximately 10\% during COP.

Let us recall that the comparisons presented in this subsection were made relative to the CubeSat opnav reference scenario analyzed at the beginning of the \textit{Results} section, highlighting the substantial improvements achieved with the additional images of Dimorphos captured by Milani and Juventas. In conclusion, as anticipated, these additional images are particularly valuable in scenarios where the system exhibits chaotic or tumbling behavior. Moreover, the CubeSats' scientific images of Dimorphos are not only crucial for advancing our scientific understanding but also play a significant role in enhancing orbit determination. This refined system knowledge is essential for the experimental phase and for ensuring the successful landing of Milani and Juventas on the asteroids.

\section{Conclusion}
This study highlights the significant contributions of the CubeSats Juventas and Milani to the Hera mission's scientific and gravity science objectives. By integrating their optical navigation data into the orbit determination process, substantial improvements are achieved in estimating key parameters such as asteroid masses, moments of inertia, rotational states, and orbital dynamics. Notably, CubeSats' opnav data prove essential for enhancing parameter estimation in both non-chaotic and chaotic Didymos-Dimorphos dynamical scenarios. The CubeSats' position estimates benefit considerably, with improvements of up to three orders of magnitude for Milani and one order for Juventas.

Additional imaging of Dimorphos within the CubeSats scientific investigation, particularly under chaotic or tumbling scenarios, is especially beneficial for both scientific investigations and mission operations. The contributions of Milani and Juventas significantly refine our understanding of Dimorphos' dynamical state and surface features while enhancing orbit determination accuracy. This improved knowledge is vital for mission planning, particularly during experimental phases and the safe landing of the CubeSats on the asteroids.

These findings underscore the critical role of CubeSats images in augmenting the capabilities of deep-space missions at small bodies. They enable more precise scientific investigations and inform the design of future optical navigation strategies. The integration of such measurements improves the triangulation of all spacecraft involved in the mission, as well as the binary asteroid system itself.

\section{Acknowledgment}
EG, PT, RLM, and MZ wish to acknowledge Caltech and the NASA Jet Propulsion Laboratory for granting the University of Bologna a license to an executable version of MONTE Project Edition S/W. EG, PT, RLM, and MZ are grateful to the Italian Space Agency (ASI) for financial support through Agreement No. 2022-8-HH.0 in the context of ESA's Hera mission. OK and BR acknowledge funding by the BELgian federal Science Policy Office (BELSPO) via the PROgramme for Development of scientific Experiments (PRODEX) by the European Space Agency (ESA).
\section{Appendix}
\begin{figure}[H]
    \centering
    \begin{subfigure}
        \centering
        \includegraphics[width=0.7\textwidth]{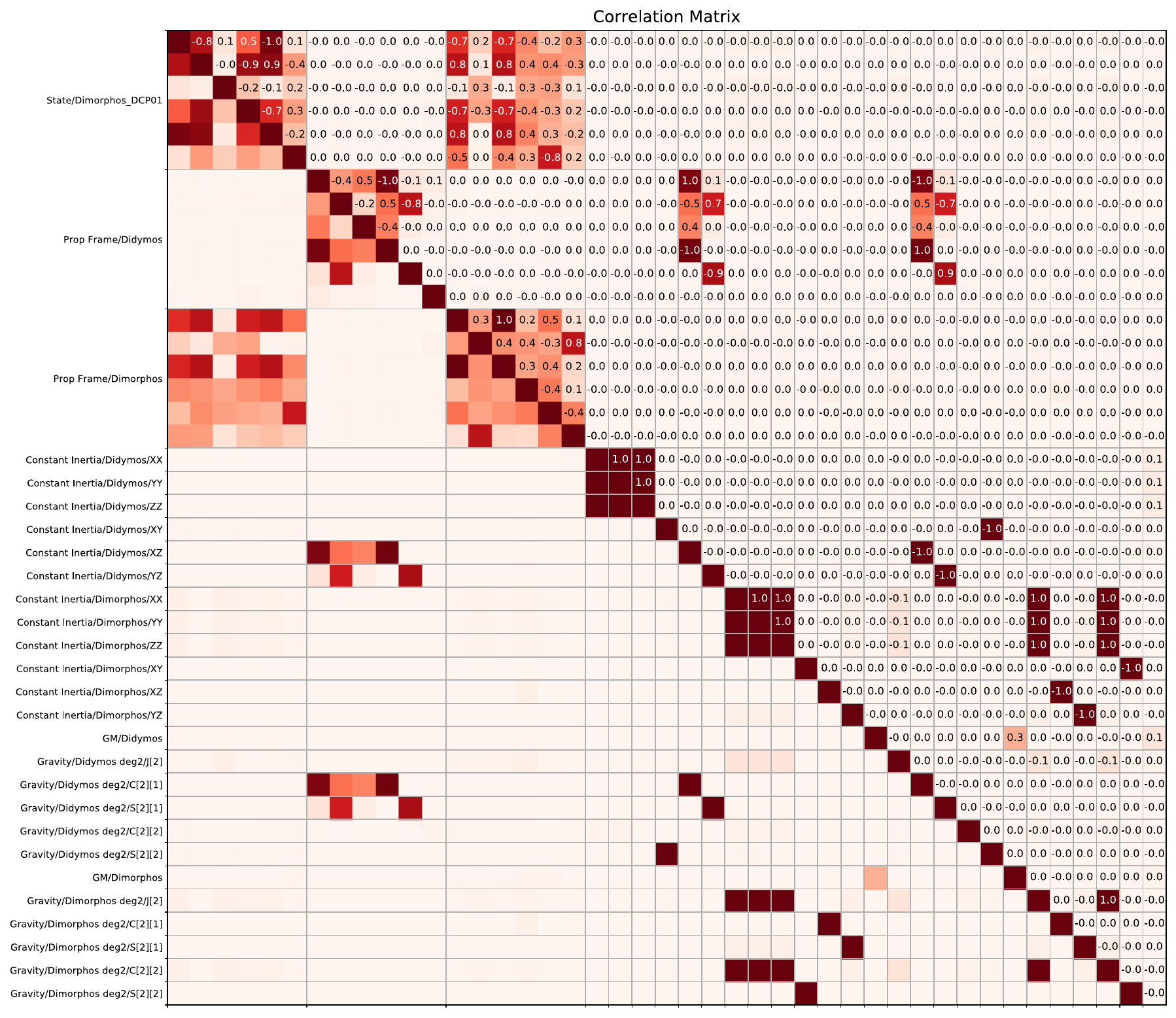} 
    \end{subfigure}
    
    \vspace{2mm} 
    
    \begin{subfigure}
        \centering
        \includegraphics[width=0.7\textwidth]{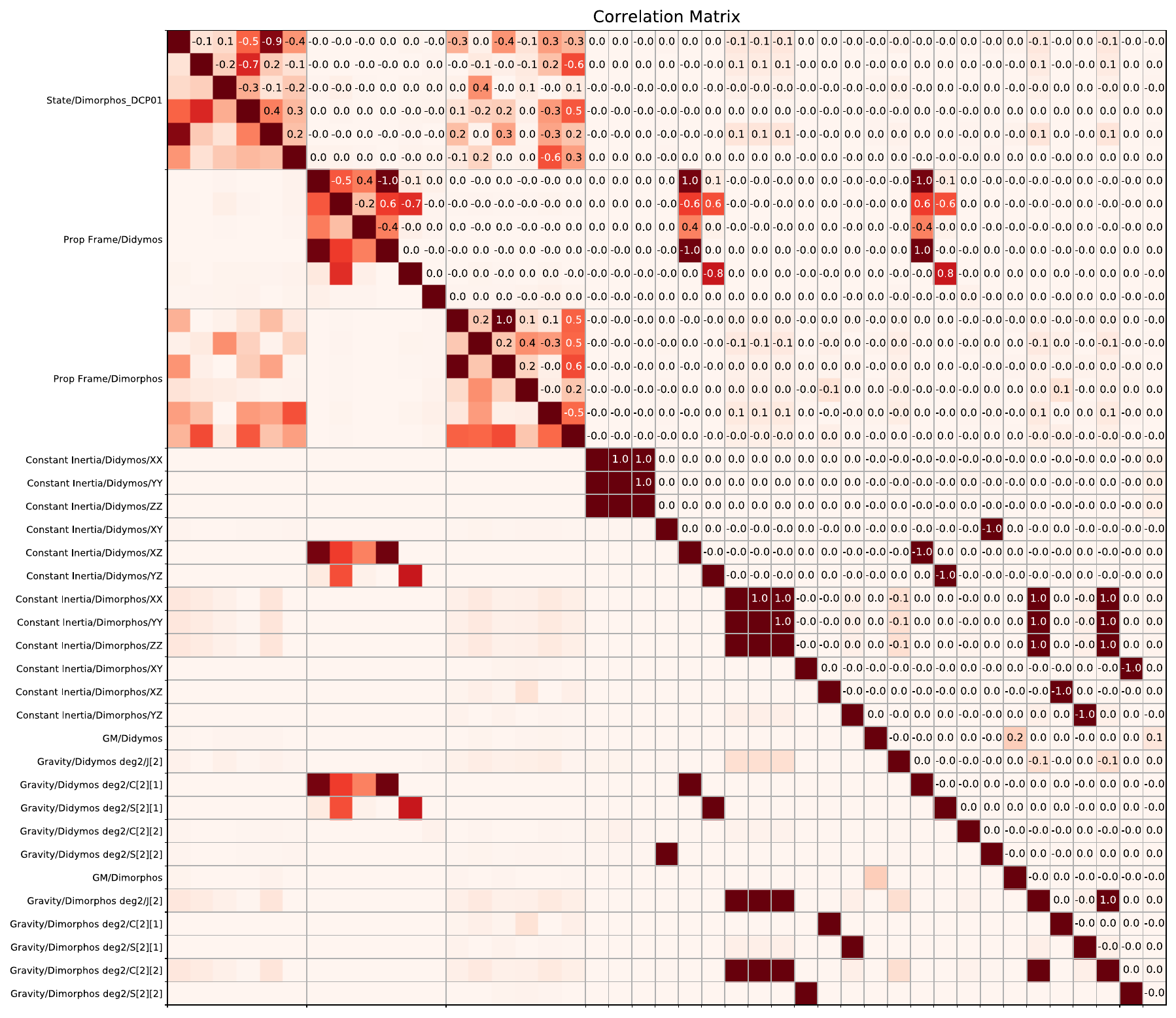} 
    \end{subfigure}   
    \caption{Correlation matrices from the F2BP chaotic simulation for key parameters of interest. The local parameters (Dimorphos' state and propagated frames), for brevity, are provided only for the first DCP radio science arc. Top panel: nominal scenario (without CubeSats images); Bottom panel: reference scenario (nominal+ CubeSats opnav). For clarity, some parameters in the correlation matrices are grouped: the state of Dimorphos parameters includes [X, Y, Z, DX, DY, DZ], the F2BP propagated frame comprises [RA, DEC, W, $\omega_X$, $\omega_Y$, $\omega_Z$]}
    \label{figscorrmatrixes}
\end{figure}

Within this appendix, we include Figure \ref{figscorrmatrixes}, which presents the covariance matrices for the chaotic dynamical scenario in the nominal simulation and the scenario incorporating CubeSats images. For brevity, only the first radio science arc of the DCP is reported herein for the local parameters (state of Dimorphos and propagated frames). The reported parameters include Dimorphos' integrated state, the integrated frames of Didymos and Dimorphos, their masses, MOI, and gravity field parameters up to degree 2.

\bibliographystyle{AAS_publication}   
\bibliography{references}   

@article{michel2022hera,
  title={The ESA Hera Mission: Detailed Characterization of the DART Impact Outcome and of the Binary Asteroid (65803) Didymos},
  author={Michel, Patrick and others},
  journal={The Planetary Science Journal},
  volume={3},
  number={7},
  pages={160},
  year={2022},
  publisher={IOP Publishing}
}

@article{cheng2018aida,
  title={AIDA DART asteroid deflection test: Planetary defense and science objectives},
  author={Cheng, Andrew F. and others},
  journal={Planetary and Space Science},
  volume={157},
  pages={104--115},
  year={2018},
  publisher={Elsevier}
}

@book{milani2010,
	author = {Milani, Andrea and Gronchi, Giovanni},
	title = {Theory of Orbit Determination},
	publisher = {Cambridge University Press},
	year = {2010}
}

@inproceedings{gramigna2024heraAAS,
    author={Gramigna, E. and Manghi, R. L. and Zannoni, M. and Tortora, P. and Gaitanas, M. and Gkolias, I. and Tsiganis, K. and and Park, R. S. and Meyer, A. and Pravec, P.},	
    title={Hera Orbit Determination Covariance Analysis of Didymos Binary Asteroid with the Full Two-Body Problem},
	booktitle={2024 AAS/AIAA Astrodynamics Specialist Conference (accepted)},
	year={2024}
}

@article{goldberg2019juventas,
	title={The Juventas CubeSat in support of ESA's Hera mission to the asteroid Didymos},
	author={Goldberg, H. R. and Karatekin, {\"O}. and Ritter, Birgit and Herique, Alain and Tortora, Paolo and Prioroc, Chiara and others},
	year={2019},
}

@article{ferrari2021trajectory,
	title={Trajectory Options for Hera’s Milani CubeSat Around (65803) Didymos},
	author={Ferrari, F. and others},
	journal={The Journal of the Astronautical Sciences},
	year={2021},
	volume={68.4},
	pages={973-994},
}

@article{ferrari2021preliminary,
	title={Preliminary mission profile of Hera’s Milani CubeSat},
	author={Ferrari, F. and others},
	journal={Advances in Space Research},
	year={2021},
	volume={67.6},
	pages={2010-2029},
}

@inproceedings{karatekin2021juventas,
	title={Juventas Cubesat for the Hera mission},
	author={Karatekin, {\"O}. and Le Bras, E. and Herique, A. and Tortora, P. and Ritter, B. and Scoubeau, M. and Moreno, V. M.},
	booktitle={European Planetary Science Congress},
	pages={EPSC2021-750},
	month={September},
	year={2021},
}

@book{bierman1977factorization,
	author = {Bierman, G. J.},
	title = {Factorization Methods for Discrete Sequential Estimation},
	series = {Mathematics in Science and Engineering},
	volume = {128},
	publisher = {Academic Press},
	address = {New York},
	year = {1977},
	pages = {1--241},
}

@article{cheng2023momentum,
  title={Momentum transfer from the dart mission kinetic impact on asteroid dimorphos},
  author={Cheng, A. F. and Agrusa, H. F. and Barbee, B. W. and Meyer, A. J. and Farnham, T. L. and Raducan, S. D. and Zanotti, G. and others},
  journal={Nature},
  volume={616},
  number={7957},
  pages={457--460},
  year={2023},
  publisher={Nature Publishing Group}
}

@inproceedings{gramigna2022hera,
  title={Hera Inter-Satellite link Doppler characterization for Didymos Gravity Science experiments},
  author={Gramigna, Edoardo and others},
  booktitle={2022 IEEE 9th International Workshop on Metrology for AeroSpace (MetroAeroSpace)},
  pages={1--5},
  year={2022},
  organization={IEEE}
}

@article{evans2018monte,
  title={MONTE: the next generation of mission design and navigation software},
  author={Evans, Scott and others},
  journal={CEAS Space Journal},
  volume={10},
  number={1},
  pages={79--86},
  year={2018},
  publisher={Springer}
}

@article{gramigna2024hera,
  title={The hera radio science experiment at didymos},
  author={Gramigna, Edoardo and Manghi, Riccardo Lasagni and Zannoni, Marco and Tortora, Paolo and Park, Ryan S and Tommei, Giacomo and Le Maistre, S{\'e}bastien and Michel, Patrick and Castellini, Francesco and Kueppers, Michael},
  journal={Planetary and Space Science},
  volume={246},
  pages={105906},
  year={2024},
  publisher={Elsevier}
}

@article{pravec2024rotational,
  title={Rotational lightcurves of Dimorphos and constraints on its post-DART impact spin state},
  author={Pravec, P and Meyer, AJ and Scheirich, P and Scheeres, DJ and Benson, CJ and Agrusa, HF},
  journal={Icarus},
  volume={418},
  pages={116138},
  year={2024},
  publisher={Elsevier}
}

@article{Poggiali2022,
	doi = {10.3847/PSJ/AC76C4},
	issn = {2632-3338},
	issue = {7},
	journal = {The Planetary Science Journal},
	month = {7},
	pages = {161},
	publisher = {IOP Publishing},
	title = {Expected Investigation of the (65803) Didymos–Dimorphos System Using the RGB Spectrophotometry Data Set from the LICIACube Unit Key Explorer (LUKE) Wide-angle Camera},
	volume = {3},
	url = {https://iopscience.iop.org/article/10.3847/PSJ/ac76c4 https://iopscience.iop.org/article/10.3847/PSJ/ac76c4/meta},
	year = {2022},
}

@article{Pugliatti2023,
	author = {M. Pugliatti and F. Piccolo and A. Rizza and V. Franzese and F. Topputo},
	doi = {10.1016/J.ACTAASTRO.2023.04.047},
	issn = {0094-5765},
	journal = {Acta Astronautica},
	month = {9},
	pages = {14-28},
	publisher = {Pergamon},
	title = {The vision-based guidance, navigation, and control system of Hera’s Milani CubeSat},
	volume = {210},
	year = {2023},
}

\end{document}